\documentclass[usegraphicx,useAMS,usenatbib]{mn2e}

\newcommand{\kms}{\,km\,s$^{-1}$}

\newcommand{\be}{\begin{equation}}
\newcommand{\ee}{\end{equation}}
\newcommand{\bd}{\begin{displaymath}}
\newcommand{\ed}{\end{displaymath}}

\title[Age spreads in the ONC]
  {Using rotation rates to probe age spreads in the Orion Nebula
  Cluster}
\author[R.D. Jeffries]
  {R.D.~Jeffries\\
  Astrophysics Group, Research Institute for the Environment, Physical
  Sciences and Applied Mathematics, Keele University, \\ Keele, 
      Staffordshire ST5 5BG
}
\date{Submitted July 9 2007}

\pagerange{\pageref{firstpage}--\pageref{lastpage}} \pubyear{2007}

\def\LaTeX{L\kern-.36em\raise.3ex\hbox{a}\kern-.15em
    T\kern-.1667em\lower.7ex\hbox{E}\kern-.125emX}

\begin{document}

\label{firstpage}

\maketitle

\begin{abstract}
The radii of young pre-main-sequence (PMS) stars in the Orion Nebula
Cluster (ONC) have been estimated using their rotation periods and
projected equatorial velocities. Stars at a given effective temperature
have a spread in their geometrically estimated projected radii that is larger
than can be accounted for with a coeval model, observational
uncertainties and randomly oriented rotation axes. It is shown that the
required dispersion in radius (a factor of 2--3 full width half maximum) can be
modelled in terms of a spread in stellar ages larger than
the median age of the cluster, although the detailed star formation
history cannot be uniquely determined using present data.  This
technique is relatively free from systematic uncertainties (binarity,
extinction, variability, distance) that have hampered previous studies of the ONC
star formation history using the conventional Hertzsprung-Russell
diagram. However, the current ONC rotational data are biased against
low luminosity objects, so the deduced dispersions in radius and
inferred age are probably underestimates. In particular, the ages of a
tail of PMS stars that appear to be $\geq 10$\,Myr old in the
Hertzsprung-Russell diagram cannot be verified with present data. If
projected equatorial velocities were measured for these objects it
could easily be checked whether their radii are correspondingly smaller
than the bulk of the ONC population.
\end{abstract}


\section{Introduction}

Whether star formation takes place rapidly on dynamical timescales or
is a quasi-static process in which protostellar cores take
many free-fall timescales to contract is one of the important debates
in theoretical star formation (e.g. Shu, Adams \& Lizano 1987;
Hartmann, Ballesteros-Paredes \& Bergin 2001; Tassis \& Mouschovias
2004; V\'azquez-Semadeni et al. 2005; Tan, Krumholz \& McKee 2006).

A key observational constraint on star formation timescales is the
possibility of significant age spreads in young clusters and
associations.  Palla \& Stahler (1999, 2000) have used ages estimated
from the Hertzsprung-Russell (H-R) diagrams of several clusters to
argue that star formation takes place over more than 10\,Myr and at an
accelerating rate towards the present day. Huff \& Stahler (2006) show
that this formation history appears to be similar in the inner and
outer parts of the Orion Nebula cluster (ONC), suggesting that the
acceleration was triggered by a global contraction of the parent cloud.
However, these conclusions are not universally accepted. Hartmann
(2001) has shown that errors in determining the luminosities of young
stars, due to binarism, uncertain extinction, variability and accretion
can lead to significant overestimates of the age spread in an H-R
diagram and obscure the detailed star formation history.

In this paper I explore an alternative technique for estimating the
extent of luminosity and (inferred) age spreads in star forming regions
using pre-main-sequence (PMS) stars in the ONC as an example.
Rotational periods and projected equatorial velocities can be combined
to give a geometric estimate of the stellar radius multiplied by the
sine of an unknown rotation axis inclination, $R\sin i$. Assuming
random axial orientations and using PMS evolutionary models, the
distributions of $R\sin i$ can be modelled in terms of a dispersion in
true radius and inferred age.  This idea was stimulated by the work of
Rhode, Herbst \& Mathieu (2001), who calculated mean $R\sin i$ as a
function of position in the H-R diagram, finding 3-sigma evidence that
stars with lower luminosities did indeed have smaller radii. Here I
provide a more sophisticated analysis of a larger dataset. This work
follows closely on from the related analysis presented in Jeffries (2007),
where a similar approach was used to provide a revised distance
estimate for the ONC.

The data base of rotational measurements used in this paper is
described in section~2. Section~3 presents the analysis methods used in
this work.  The results are presented in section~4 and discussed in
section~5.

\section{The observational database}

\label{observations}

\begin{figure}
\includegraphics[width=84mm]{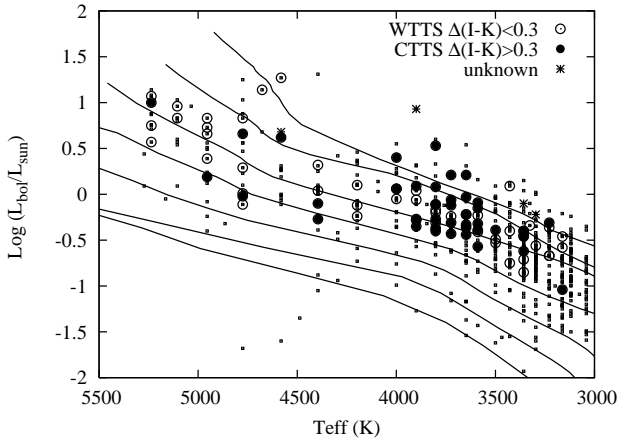}
\caption{A Hertzsprung-Russell diagram for the 95 stars of the ONC
  rotation sample assuming an ONC distance of 392\,pc (see text). Solid
  symbols indicate objects classified as accreting T-Tauri stars on the
  basis of an $I-K$ colour excess greater than 0.3 mag. Open
  symbols indicate non-accreting objects using the same
  criterion. Small symbols indicate all proper-motion selected members of
  the ONC from Hillenbrand (1997). The lines on the plot are isochrones
  from the models of D'Antona \& Mazzitelli (1997). From the top
  downwards these isochrones are at 0.1, 0.3, 1.0, 3.0, 10, 30 and
  100\,Myr respectively.
}
\label{hrdiag}
\end{figure}

\begin{figure*}
\centering
\begin{minipage}[t]{0.45\textwidth}
\includegraphics[width=80mm]{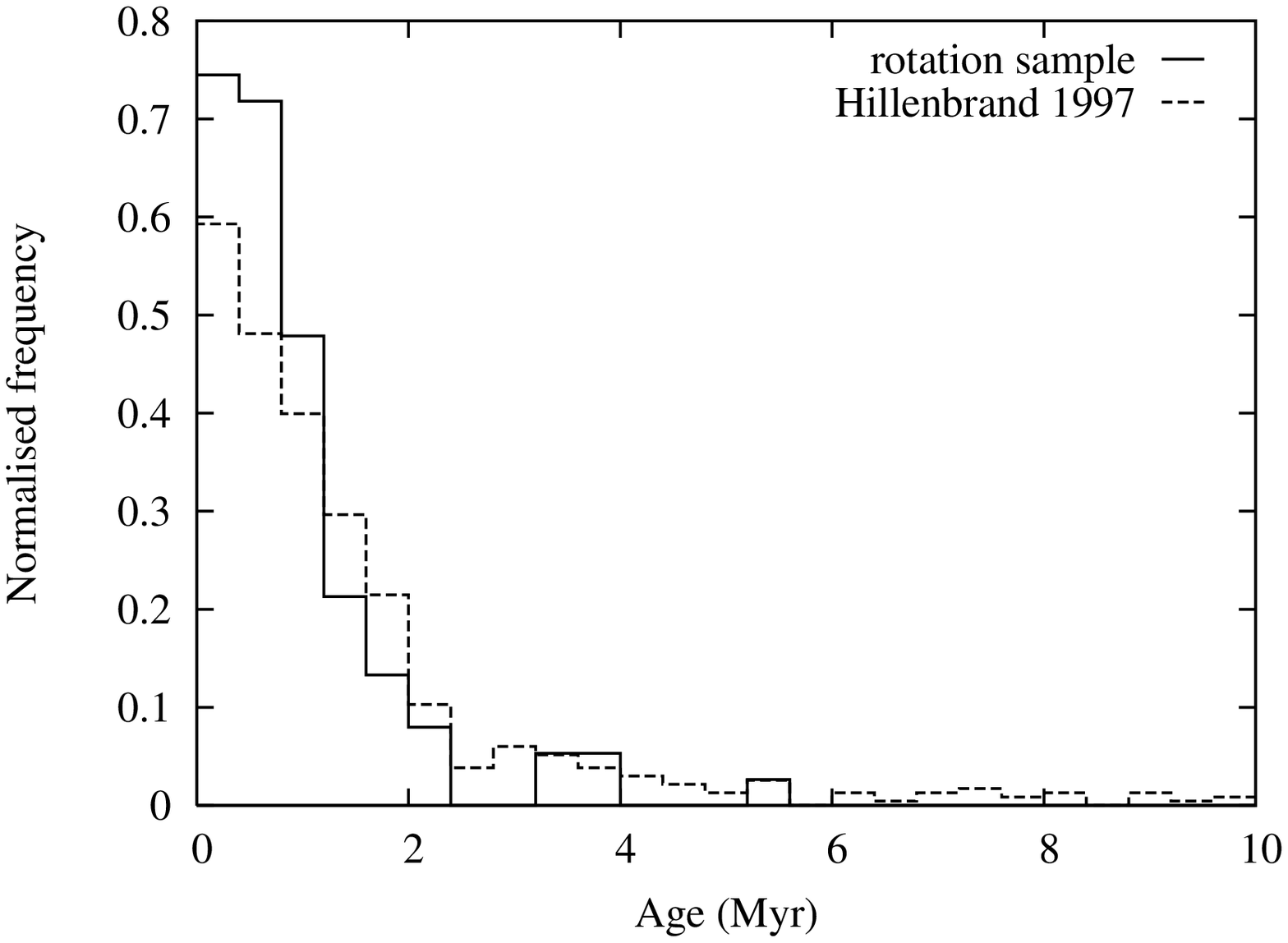}
\end{minipage}
\begin{minipage}[t]{0.45\textwidth}
\includegraphics[width=80mm]{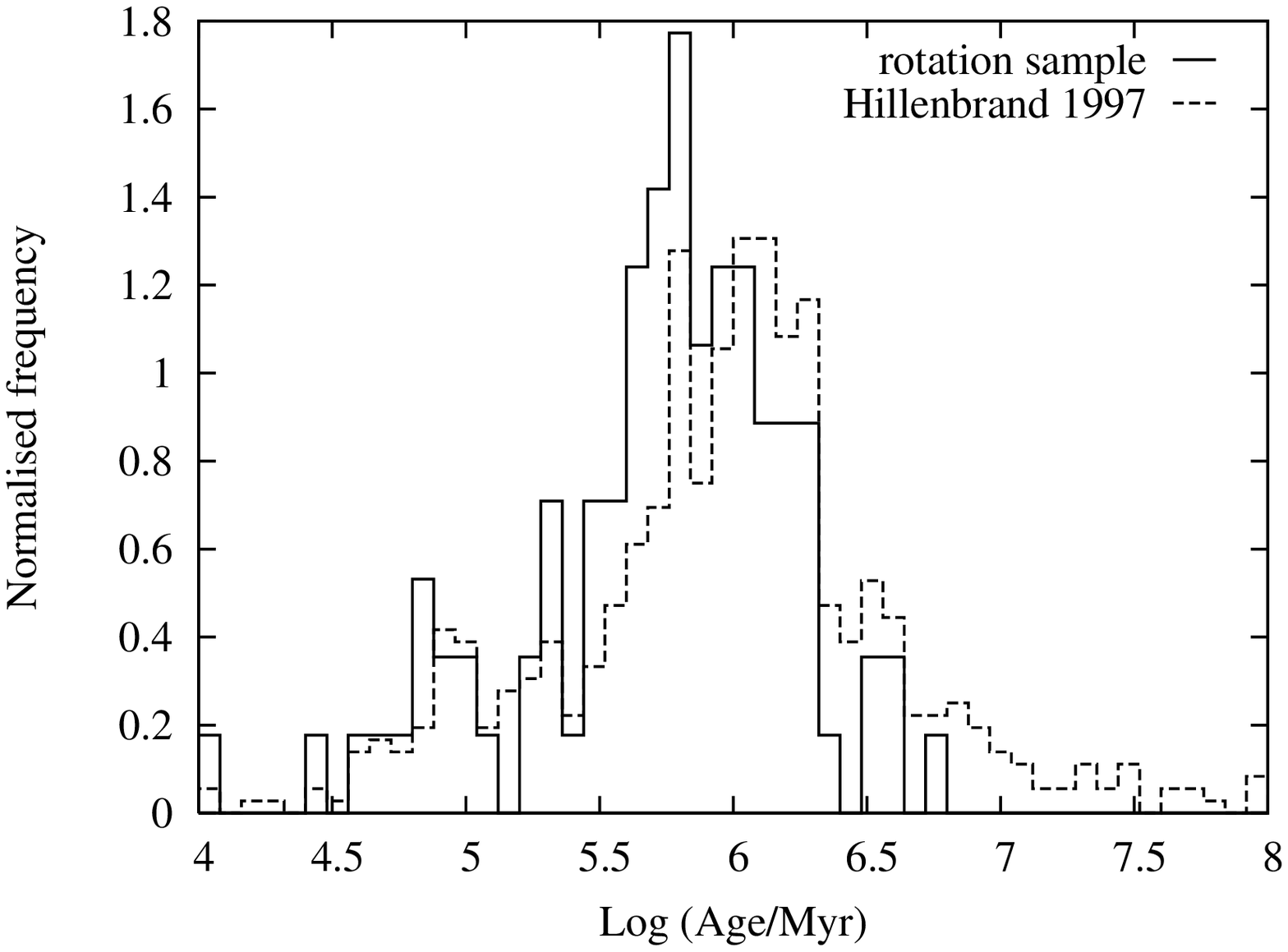}
\end{minipage}
\caption{Deduced age distributions from Fig.~\ref{hrdiag} for the
  rotation sample and for proper-motion selected ONC members from
  Hillenbrand (1997). The left panel is binned linearly showing the
  ``exponentially accelerating'' star formation rate proposed by Palla
  \& Stahler (1999). The right panel is binned logarithmically. It is
  clear from this plot that a tail of older stars that is present among
  the proper-motion selected ONC members is not present in the
  rotation sample.}
\label{age}
\end{figure*}

\begin{figure*}
\centering
\begin{minipage}[t]{0.45\textwidth}
\includegraphics[width=80mm]{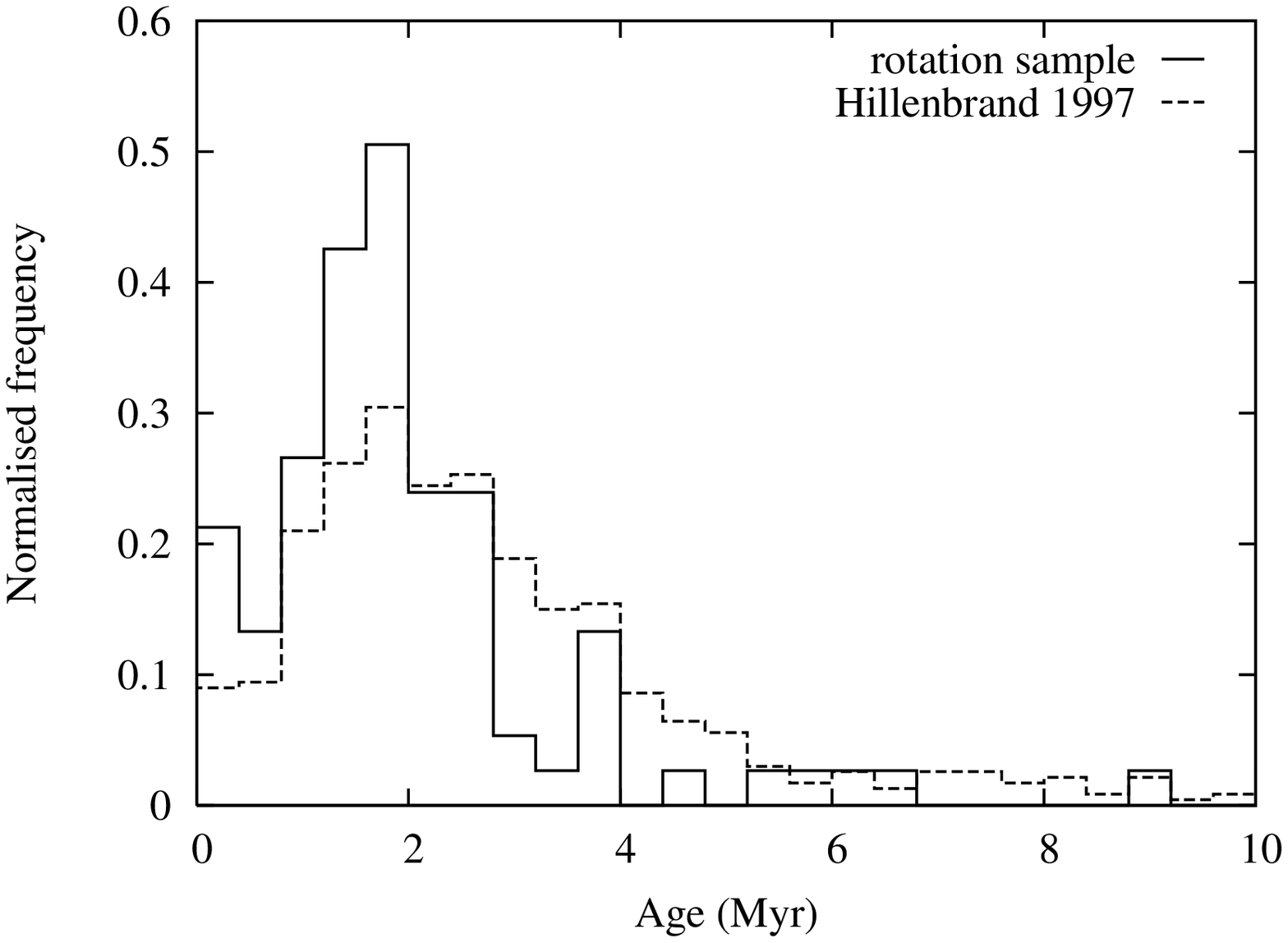}
\end{minipage}
\begin{minipage}[t]{0.45\textwidth}
\includegraphics[width=80mm]{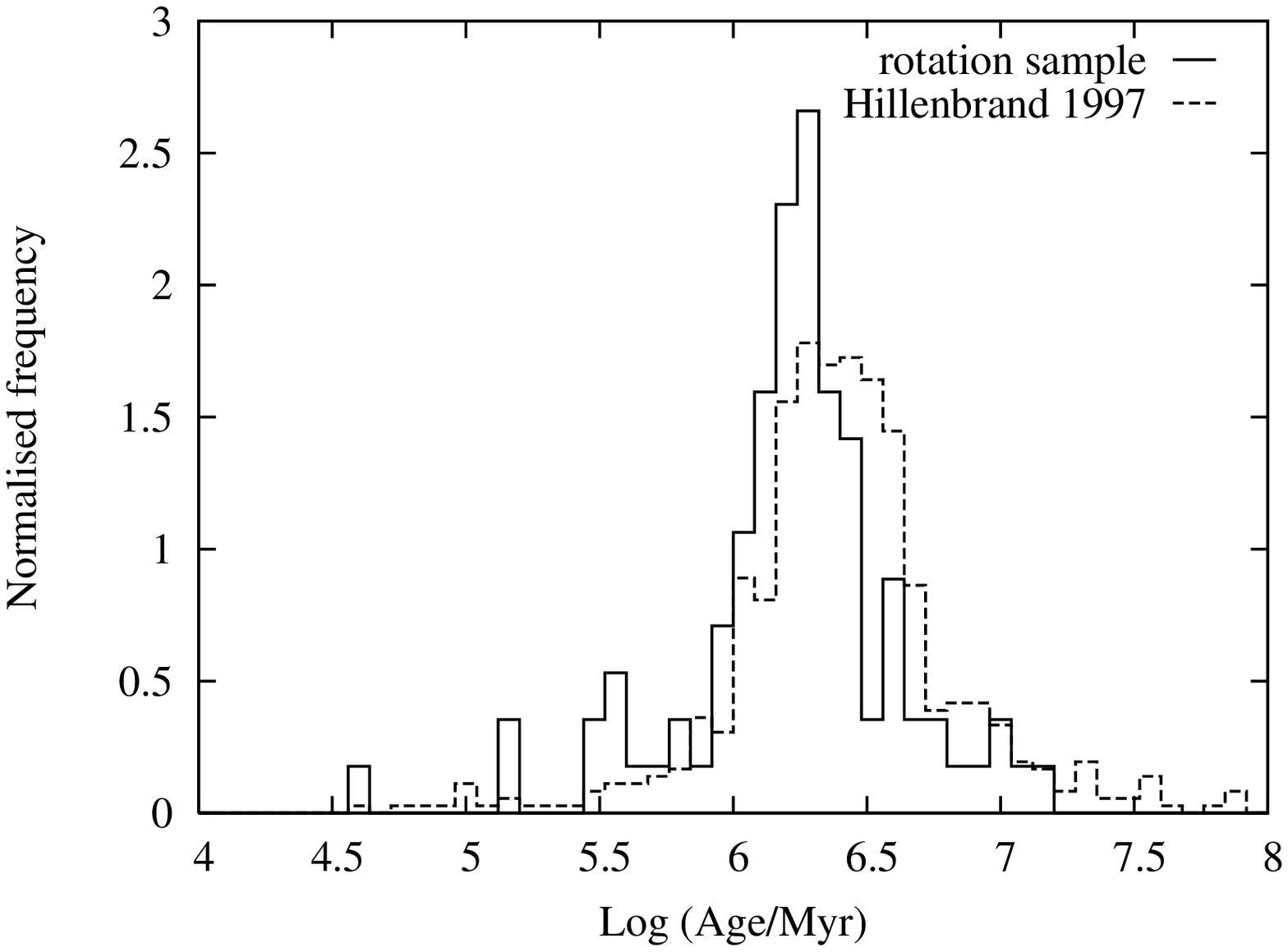}
\end{minipage}
\caption{Deduced age distributions from the H-R diagram for the
  rotation sample and for proper-motion selected ONC members, but this
  time using the Siess et al. (2000) models to assign ages. Left and
  right panels show linearly and logarithmically binned ages.
}
\label{age2}
\end{figure*}

The ONC is among the best studied star forming regions and the premier
cluster for investigating star formation and early stellar evolution.
It is relatively nearby, very young ($392\pm 34$\,pc, $\simeq$1--2\,Myr
-- Jeffries 2007; $389^{+24}_{-21}$\,pc -- Sandstrom et al. 2007) and
contains a large population of stars and brown dwarfs covering the
entire (sub)stellar mass spectrum ($0.01<M/M_{\odot}<30$ -- see
Hillenbrand 1997; Slesnick, Hillenbrand \& Carpenter 2004).

Jeffries (2007) assembled a database of 95 K- and M-type PMS stars in
the ONC, with model-dependent masses of about 0.2--2\,$M_{\odot}$ and which have
proper-motion or radial velocity membership credentials, spectral types
and luminosities (Hillenbrand 1997), effective temperatures based
on the work by Cohen \& Kuhi (1979), indicators of whether objects were
actively accreting (based on $I-K$ excess, from Hillenbrand [1997]),
rotation periods (Herbst et al. 2002) and $v\sin i$ (measured by
Rhode et al. [2001] or Sicilia-Aguilar et al. [2005]).
The typical precision of the measurements are $\pm 1$ spectral subclass,
$\pm 150$\,K in effective temperature, $<1\%$ uncertainty in the periods
and $\simeq 10$\% uncertainties in $v\sin i$. The smallest $v \sin i$
values that are taken to be reliably measured are 13.6\kms\ and
10.0\kms\ for the Rhode et al. and Sicilia-Aguilar et al. measurements respectively.

There are various filters applied to this data and a number of possible
selection effects. The
most important of these with regard to possible radius or age spreads are: (i) a
bias against the inclusion of classical T-Tauri stars (CTTS), because
their rotation periods can be difficult to distinguish from other
non-periodic variability; (ii) a bias against slowly rotating stars,
either because their periods are too long to reliably detect, they
are not so magnetically active or their $v \sin i$ is too small to be
reliably measured; and 
(iii) the sample of stars which are bright enough to
obtain spectroscopy is biased towards optically brighter and hence
intrinsically more luminous (and probably younger) stars.

If CTTS are predominantly younger than non-accreting weak-lined T-Tauri
stars (WTTS) then the first effect will bias the sample towards older
ages. The second effect is more complicated. As PMS stars contract they
spin-up, but they may also lose angular momentum through winds or
star-disc interactions as they get older.  The longer period stars that
remain in our sample appear to have larger radii on average (from radii
estimated by Hillenbrand 1997) and may be younger. On the other hand,
at a given rotation period a smaller (older) star will have a lower
$v\sin i$.  Therefore these selection effects probably bias our sample
against the inclusion of some of the youngest and oldest
stars. Finally, any bias against low-luminosity stars will potentially
exclude older stars, especially in the lowest mass objects in our
sample. The discussion of completeness in Hillenbrand (1997) suggests
that any incompleteness in the stars with spectroscopy will arise
from the requirements of obtaining $v \sin i$ using high resolution
spectroscopy, which is restricted to brighter stars. Spectral types
can be obtained using low resolution spectroscopy for much fainter objects.

An illustration of these potential effects is provided by comparing the
H-R diagrams for our ``rotation sample'' and for proper-motion
selected ONC members with temperatures and luminosities from
Hillenbrand (1997). The H-R diagram is shown in Fig.~1, along with isochrones
from D'Antona \& Mazzitelli (1997). The luminosities from
Hillenbrand (1997) have been adjusted to correspond to the 392\,pc
found by Jeffries (2007), rather than the 470\,pc assumed by
Hillenbrand (1997). 

Model-dependent PMS stellar ages are estimated from these
isochrones. Histograms of age are shown in Fig.~\ref{age} for the
rotation sample and the full ONC sample, using linear and logarithmic
age bins. The median ages are 0.62\,Myr and 1.03\,Myr for the
rotation sample and the full sample respectively.
The linearly binned age distribution of the Hillenbrand (1997)
sample in Fig.~\ref{age} shows the pseudo-exponential decline
characterised as ``accelerating star formation'' by Palla \& Stahler
(1999). This translates into a peak in the number of stars per
logarithmic age interval at around 1\,Myr.
Fig.~\ref{age} also demonstrates that the tail of ``older''
($>3$\,Myr) stars appears to be absent from the rotation sample.  A
Kolmogorov-Smirnov (K-S) test suggests that there is a $<1$ per cent
probability that the two age distributions are consistent. This
selection effect arises chiefly from the bias towards
brighter targets. Some of the age bias might be thought to arise because older, smaller
objects could have unresolveable $v \sin i$ (see Rebull, Wolff \& Strom
2004). However, the H-R
diagram-based age distribution for objects which have unresolved $v
\sin i$ but otherwise match the sample selection criteria is in fact
similar to that of the rotation sample.

Stars in the rotation sample are classified as CTTS or WTTS on the
basis of whether their $I-K$ colour excess is greater or less than 0.3
mag (see Hillenbrand et al. 1998). A K-S test on the age distributions
of these two populations reveals no significant evidence for a
difference.

The age distributions in Fig.~\ref{age} are of
course model dependent. I have repeated these exercises using the
Siess, Dufour \& Forestini (2000) models (the $Z=0.02$ variety) 
and these are shown in Fig.~\ref{age2}. 
The Siess et al. models give older ages and
the evidence for ``accelerating'' star formation up to the present day
now disappears to be replaced by a peak in the star formation rate in
the recent past. However, the conclusions regarding a bias in the
rotation sample and the missing tail of older stars remain.
The median ages using the Siess et al. (2000) models are 1.55\,Myr and 2.51\,Myr
for the rotation sample and full samples respectively.

\section{Computational methods}

\subsection{Monte Carlo Models}

The background to the method used in this paper is described 
by Jeffries (2007). The projected equatorial radius of a PMS
star can be estimated from
\be
(R \sin i)_{\rm obs} = \frac{k}{2\pi}\, P_{\rm obs}\, (v \sin i)_{\rm obs} 
\label{1}
\ee
where $i$ is the inclination of the rotation axis to the line of sight,
$P_{\rm obs}$ is the observed period, $(v\sin i)_{\rm obs}$ is the
observed projected equatorial velocity and $k$ a constant dependent on
the system of units in use. For any group of stars a mean $R\sin i$
could be found and divided by an average projection factor to obtain an
estimate of the true radius. This was the approach adopted by Rhode et
al. (2001), but is subject to a number of problems. First, many
objects are not included in the considered samples because their $v\sin
i$ values are too small to be resolved or their periods could not be
measured because $\sin i$ was too small. Second, uncertainties in the measured $P$ and
$v\sin i$ values themselves mean that the underlying distribution of
projection factors is modified from a pure sinusoidal form.  The aim
here is to model the distribution of $(R \sin i)_{\rm obs}$ using a
Monte Carlo simulation, starting from an assumed distribution of
$R_{\rm true}$, the assumption of random axial inclination and taking
account of the selection effects and measurement errors.

If we accept that $P_{\rm obs}$ is
related to $P_{\rm true}$, the true rotation period in the absence of
measurement uncertainties, according to a normal error distribution
characterised by a fractional measurement uncertainty $\delta P$, then
\be
P_{\rm obs} = P_{\rm true}\, ( 1 + \delta_P U)\, ,
\ee
where $U$ is a random number drawn from a Gaussian distribution with
mean of zero and unit standard deviation. The precision of the periods
is generally very good and we assume $\delta_P=0.01$ in all cases.
Similarly, there is a
relationship between the true and observed projected
equatorial velocity
\be
(v \sin i)_{\rm obs} = v_{\rm true}\, \sin i\, (1 + \delta_v U)\, ,
\label{eqn4}
\ee
where $\delta v$ is the fractional observational uncertainty in $(v
\sin i)_{\rm obs}$ (taken from the papers where the measurement was presented). 
In this case $v_{\rm true}$ and $\sin i$ are
split into separate factors in order to properly model the
observational lower limits to $v \sin i$ and any selection effect on
the possible values of $i$ (see below).

Equations 1--3 can be combined to give
\be
(R \sin i)_{\rm obs} = R_{\rm true}\, (1 + \delta_{P}U_1)\,(1+\delta_v
    U_2)\, \sin i\, ,
\label{eqn6}
\ee
where $\sin i$ can be generated assuming random axial orientation and
$U_1$ and $U_2$ are different random numbers taken from a unit Gaussian distribution.
Note that in this formulation $(R \sin i)_{\rm obs}$ is independent of
the assumed $P_{\rm true}$ or $v_{\rm true}$ values, but that this is
not the case once selection effects are considered (see below).

For a given observational sample, $10^{4}$ randomised 
values of $(R \sin i)_{\rm obs}$ are
generated for each star.  The initial $R_{\rm true}$
distribution is taken from model isochrones at the $T_{\rm eff}$
corresponding to each star in the observed sample. As a
further refinement $R_{\rm true}$ can be drawn randomly from a
distribution, either in terms of a spread about the isochronal radius,
or specified as a distribution of age (see
section~\ref{agedist}).

There are some further complications which are dealt with in the
model. First, because the isochrones have a significant slope in the
H-R diagram, then an uncertainty in $T_{\rm eff}$ leads to
additional dispersion in the expected $R_{\rm true}$. This is accounted for with a
$\pm 150$\,K Gaussian perturbation in $T_{\rm eff}$ when $R_{\rm true}$ is
calculated from the evolutionary models.

Second, the observational sample will be incomplete for objects which
have a low inclination angle because starspot modulation may be
difficult to observe in these stars (see Jeffries 2007 for a
discussion).  This is dealt with by adopting a threshold
inclination angle $i_{\rm th}$, below which randomised trials will be
rejected from the model distribution.  This free parameter principally
affects the low end of the $R \sin i$ distribution. It is assumed that
$i_{\rm th}$ is independent of rotation period. 

There is also the issue of limits imposed by spectral resolution on the
values of $(v \sin i)_{\rm obs}$ that can be recorded. This is dealt
with by discarding randomised trials which have a $v \sin i$
lower than the threshold appropriate for the dataset in
question.  The minimum measurable value of $(v \sin i)_{\rm obs}$ is
well defined, but the rejection process requires that an
initial $v_{\rm true}$ distribution is also
specified. Here I follow the example in Jeffries (2007) and
hypothesise a simple $v_{\rm true}$ distribution, which after
multiplying by the projection factor and accounting for observational
uncertainties, provides a good match (measured with a K-S test) to the
observed $v\sin i$ distribution. The exact choice of $v_{\rm true}$
distribution has very little effect on the simulated $R\sin i$
distributions (see also Jeffries 2007).

\subsection{Age distributions}
\label{agedist}

A number of possible radius or age distributions can be tested to see
whether they provide a reasonable description of any particular sample.
The scenarios investigated are: (1) a co-eval population; 
(2) a Gaussian distribution of $\log_{10} R$ around
an isochronal locus; (3) a radius
calculated according to a Gaussian distribution in $\log_{10}$ age; (4) a
radius calculated from an age distribution which exponentially decays
beyond an initial starting age; and (5) a radius calculated according
to the age distributions inferred from the traditional H-R diagram (see
Figs.~\ref{age} and \ref{age2}). 

Because $R_{\rm true}$ is expected to vary with $T_{\rm eff}$ at a
given age then the distributions of $R_{\rm true}$ should be specified
as a function of $T_{\rm eff}$.  This problem is finessed by
normalising $R_{\rm true}$ by its value at an age of 3\,Myr and at the
$T_{\rm eff}$ of a given star (for the observations) or trial (for the
models).  This effectively collapses the two-dimensional distribution
along the isochrones, reducing the observed distribution to
one-dimensional form and will be termed the distribution of normalised projected
radii, $R\sin i/R_{\rm 3Myr}$.

The results from scenarios 1 and 2
will be largely independent of the choice of evolutionary model
as they rely only on normalising the $R_{\rm true}$ values with an
appropriate isochrone of $R$ vs $T_{\rm eff}$. However, the absolute
ages and age spreads in scenarios 3--5 could be very dependent on
choice of evolutionary model. With this in mind the evolutionary models
of D'Antona \& Mazzitelli (1997 -- hereafter DAM97) and the solar
metallicity models of Siess, Dufour \& Forestini (2000, hereafter S00)
have both been tested.

There are too many free parameters and too few data points in the
binned distribution of $(R \sin i)_{\rm obs}$ (see
section~\ref{results}) to attempt an inversion to a true radius
distribution or to perform chi-squared fitting -- although this may be
possible in future with more numerous data points. Instead, K-S tests
are used to determine whether a model is capable of providing a
satisfactory description of the observed (cumulative) $(R \sin i)_{\rm
obs}/R_{\rm 3Myr}$ distribution and to show which models can be ruled
out by the data.  In each scenario the parameter which determines the
mean $R \sin i/R_{\rm 3Myr}$ value, i.e. the central isochronal age,
has been adjusted to minimise the K-S statistic and 
maximise consistency between data and model.

The main goals are to answer the following questions:
\begin{enumerate}
\item Is the normalised $R\sin i$ distribution of the ONC rotation sample
  consistent with a coeval sample that has no dispersion in radius about an
  isochronal value?

\item If not, then what is the spread in radius about an
  isochronal value that could best explain the observed normalised $R\sin i$
  distribution?

\item If the dispersion in radius is modelled as an age distribution
  then what is the age spread implied by the data and is any
  particular form of the age spread (Gaussian, exponential) preferred?

\item Is the distribution of normalised $R \sin i$ consistent with the age
  distribution inferred from the positions of the same stars in the H-R diagram? 
\end{enumerate}

\section{Results}
\label{results}

\begin{figure*}
\centering
\begin{minipage}[t]{0.45\textwidth}
\includegraphics[width=84mm]{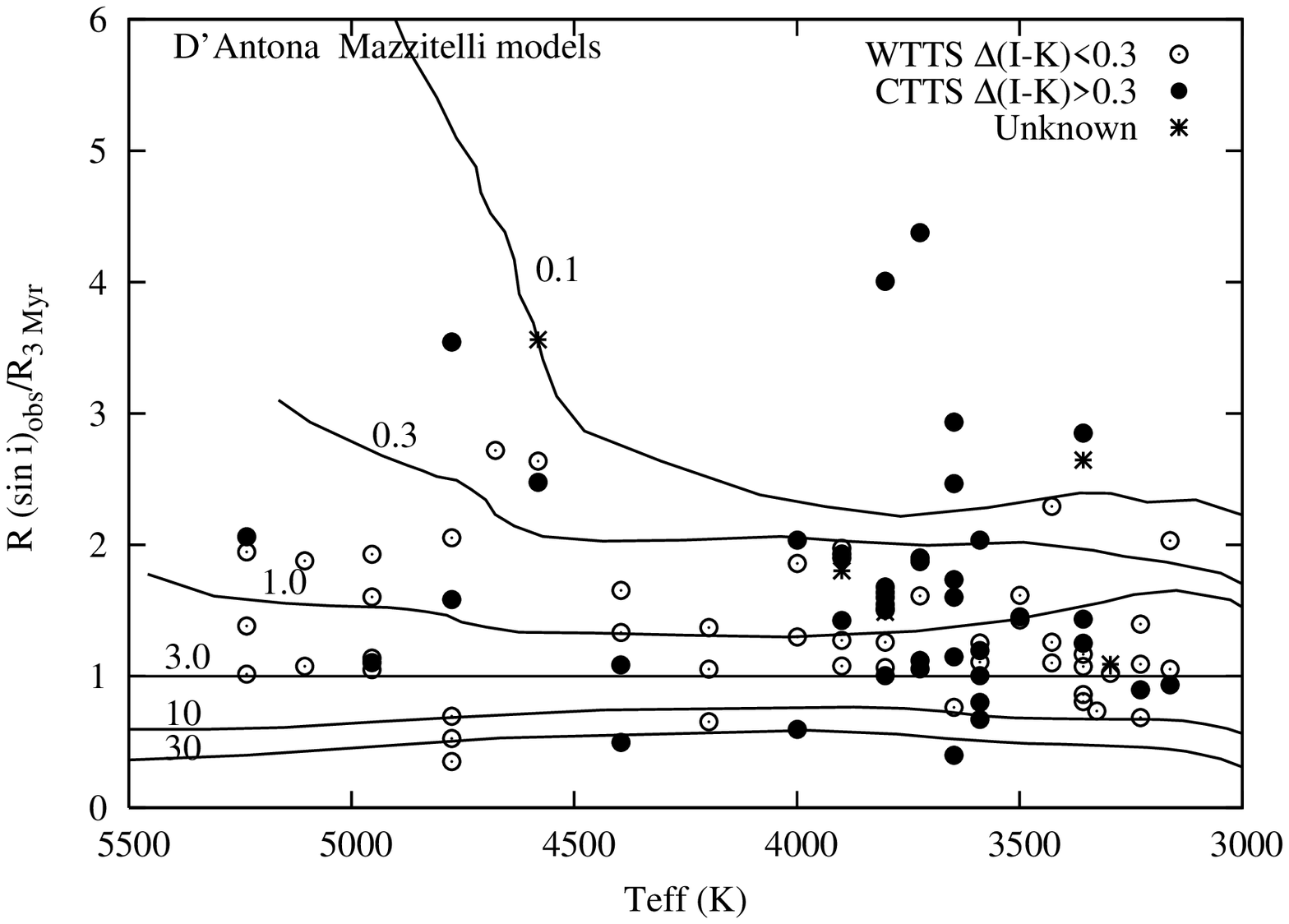}
\end{minipage}
\begin{minipage}[t]{0.45\textwidth}
\includegraphics[width=84mm]{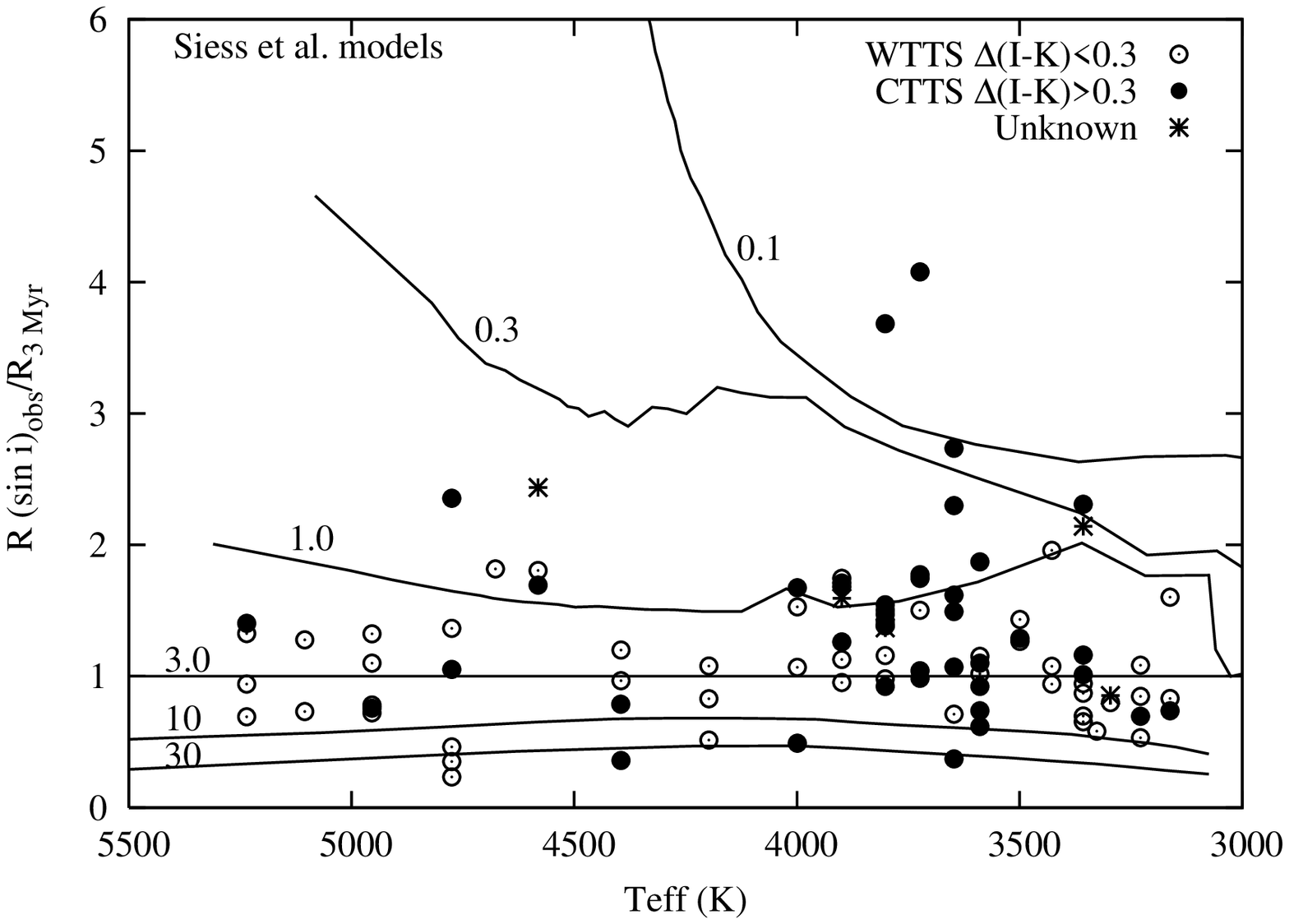}
\end{minipage}
\caption{Normalised $(R\sin i)_{\rm obs}$ versus $T_{\rm eff}$ for the
  rotation sample. $(R \sin i)_{\rm obs}$ was calculated using
  equation~\ref{1} and then divided by the isochronal value of $R$ at
  3\,Myr. The two panels show the results of these calculations using
  the evolutionary models of DAM97 (left) and S00 (right). CTTS and
  WTTS (classified according to their near-IR excesses) are given
  different symbols. In each panel the solid lines indicate isochrones
  of normalised radius at 0.1, 0.3, 1.0, 3.0, 10 and 30\,Myr.
}
\label{nrsini}
\end{figure*}

\subsection{The projected stellar radii}

Figure~\ref{nrsini} shows the normalised projected radii for the
rotation sample as a function of $T_{\rm eff}$. Both the DAM97 and S00
models have been used to normalise the $(R\sin i)_{\rm obs}$ values
calculated from equation~\ref{1} and the results are shown in the left
and right hand panels respectively, compared to isochrones of normalised
radius from the same models. In neither case is there any evidence for
a trend in $(R\sin i)_{\rm obs}/R_{\rm 3Myr}$ with $T_{\rm eff}$ (a
least squares fit gives a flat line within 1-sigma uncertainty in both
instances), which supports the approach of using the distribution of
$(R\sin i)_{\rm obs}/R_{\rm 3Myr}$ as a $T_{\rm eff}$-independent
indicator of the true spread in radii (or age). The mean values of
$(R\sin i)_{\rm obs}/R_{\rm 3Myr}$ are $1.51\pm 0.08$ and $1.24\pm
0.07$ for the DAM97 and S00 models respectively. Hence
the mean age of the ONC (as judged by the stellar radii) is
significantly younger according to the DAM97 models than according to the
S00 models, but in both cases is $<3$\,Myr because $(R\sin i)_{\rm obs}$
should be a lower limit to $R_{\rm true}$.

\subsection{A coeval population}

\begin{figure*}
\centering
\begin{minipage}[t]{0.45\textwidth}
\includegraphics[width=80mm]{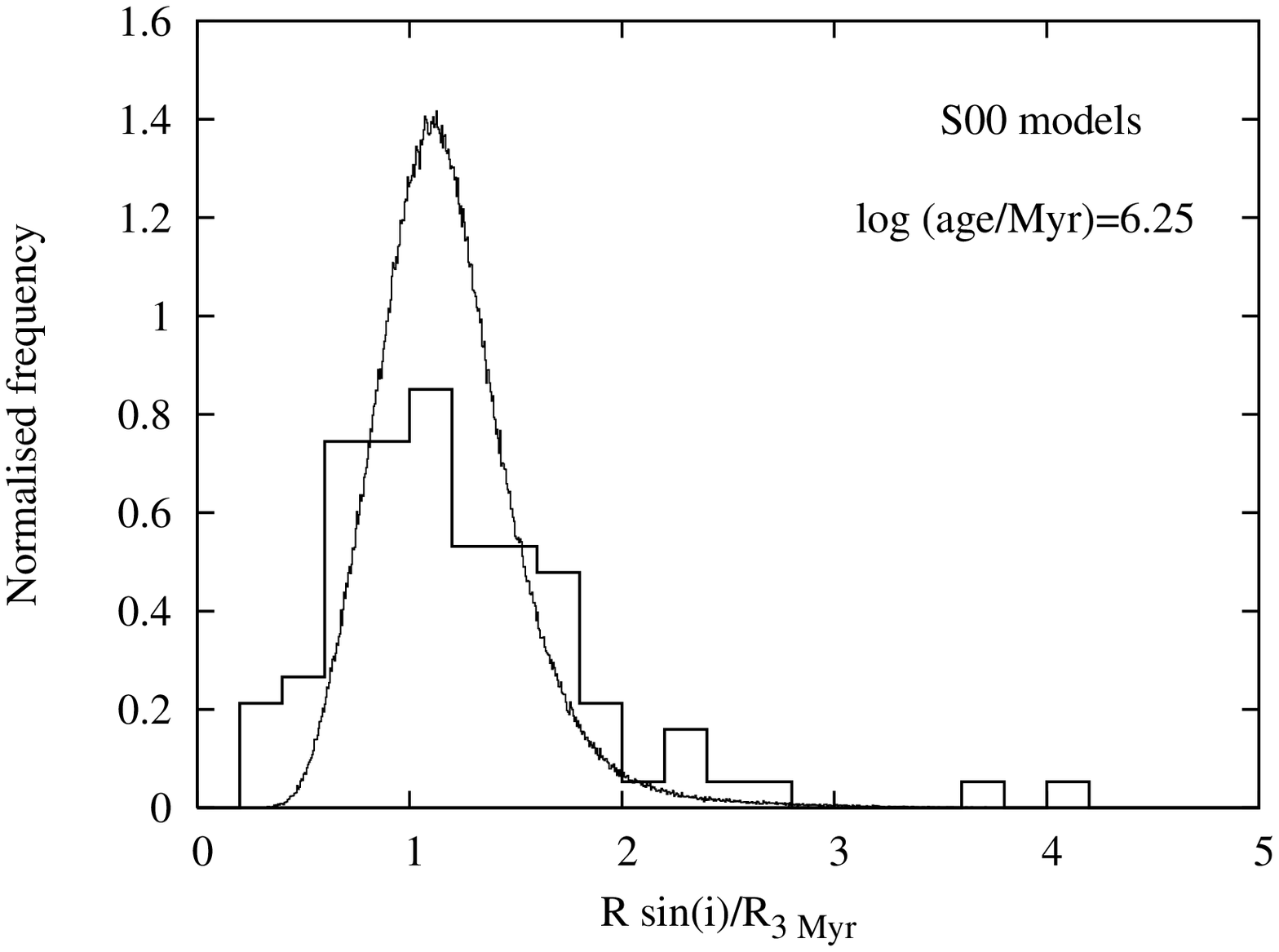}
\includegraphics[width=80mm]{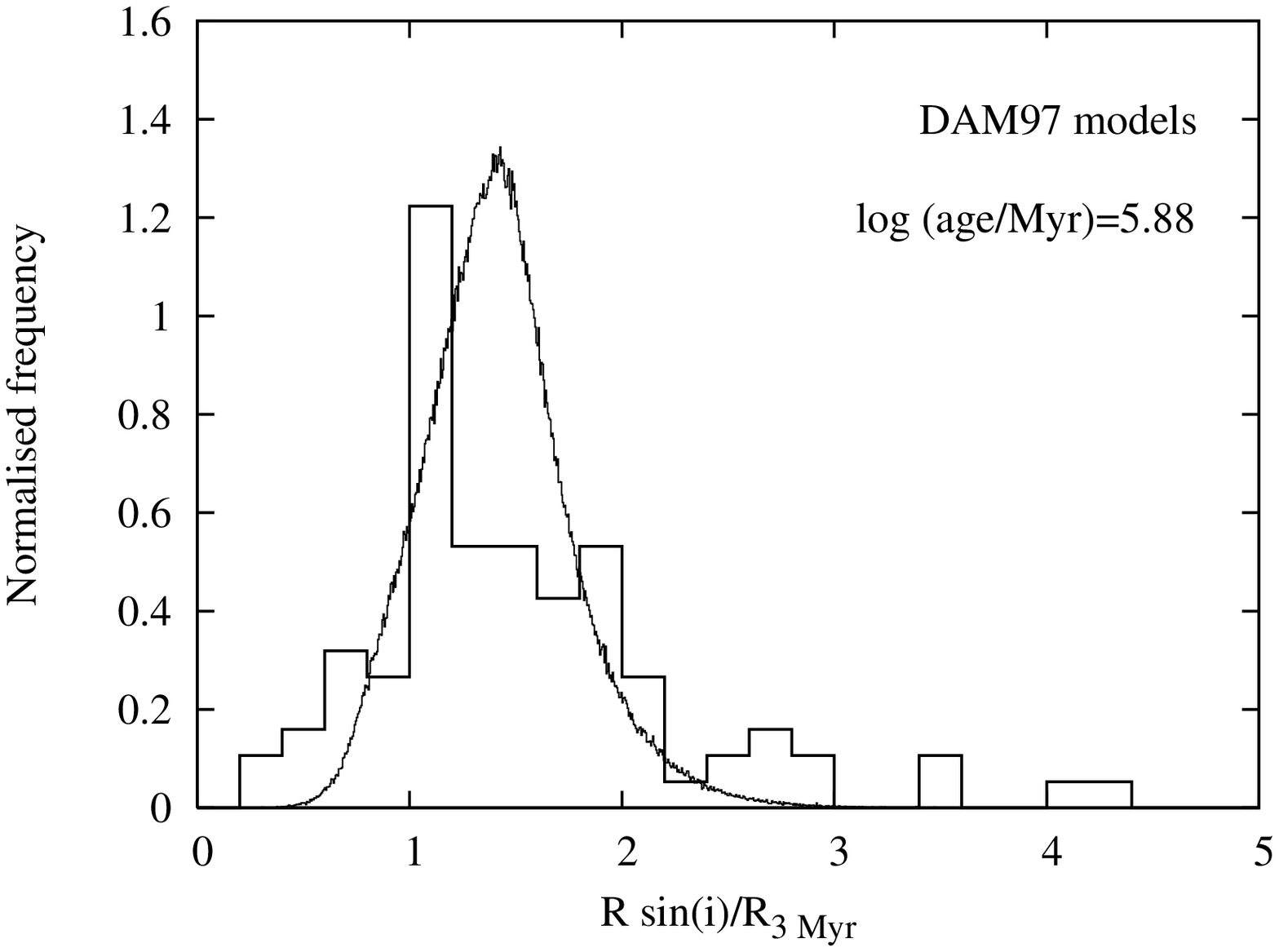}
\end{minipage}
\begin{minipage}[t]{0.45\textwidth}
\includegraphics[width=80mm]{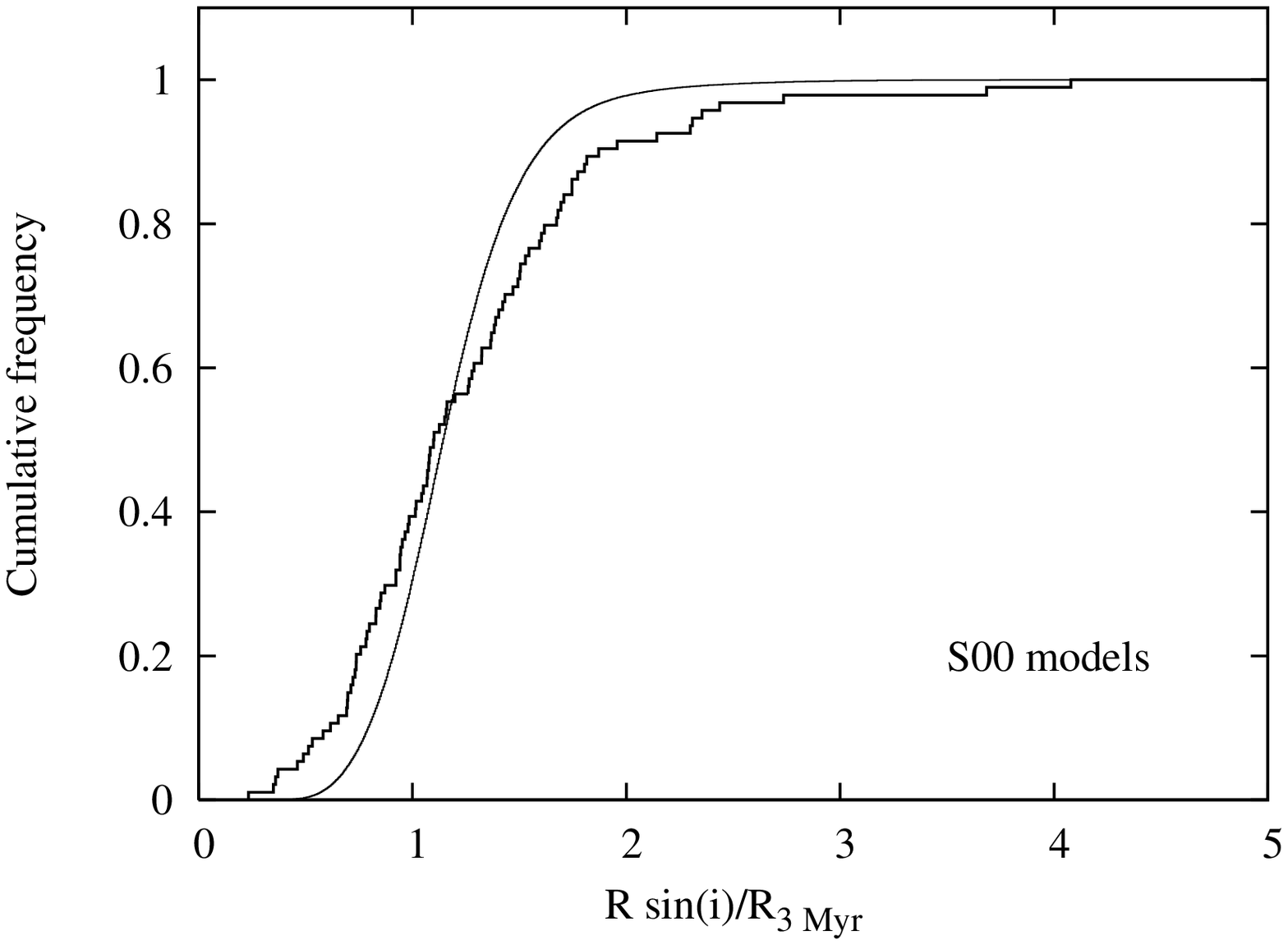}
\includegraphics[width=80mm]{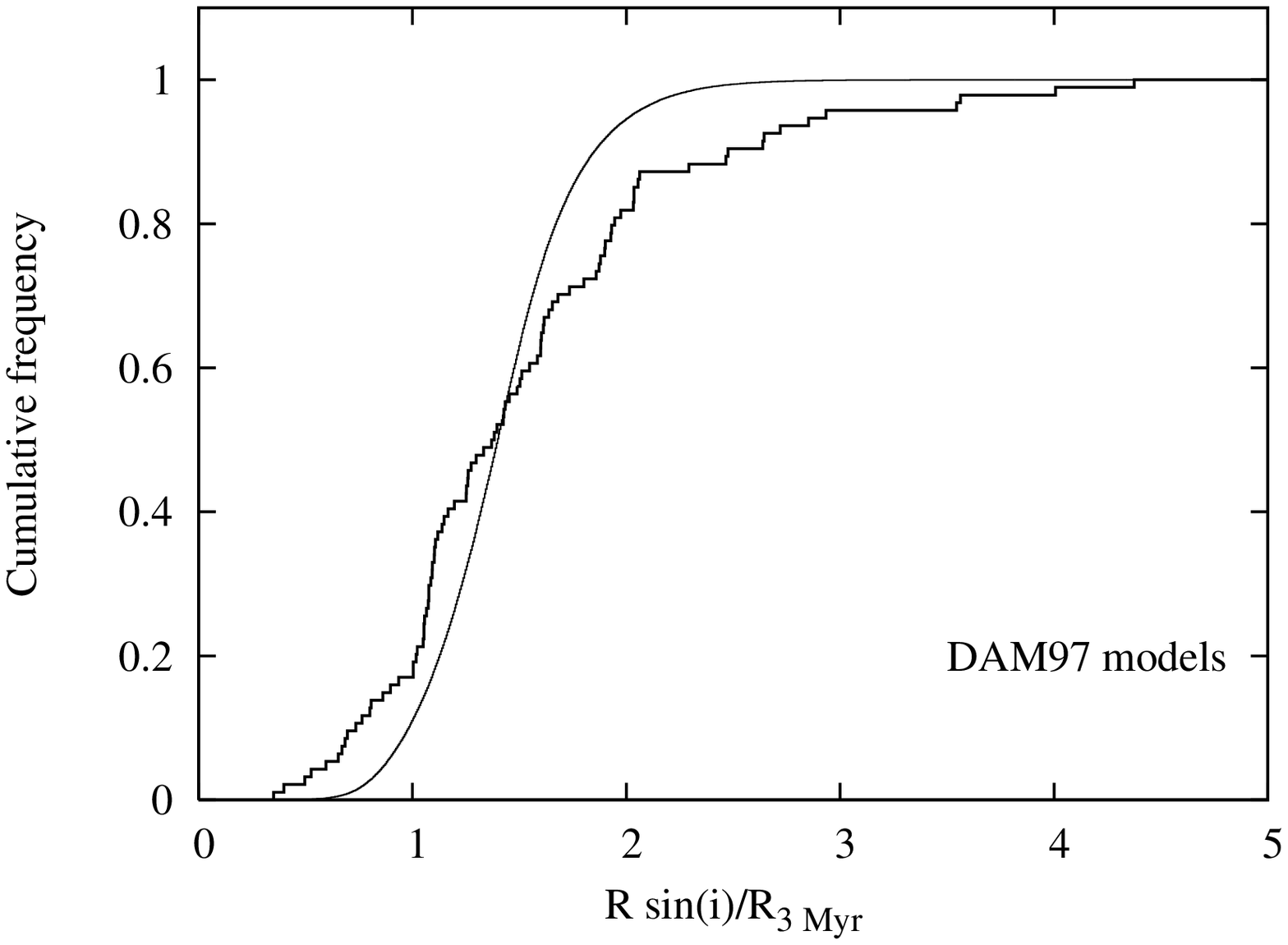}
\end{minipage}
\caption{(Left) Normalised distributions of $(R\sin i)_{\rm obs}/R_{\rm
  3 Myr}$ (binned data) compared with models which assume a coeval
  population. The values of $R_{\rm 3Myr}$ are taken
  from the models of Siess et al. (2000, top) or D'Antona \& Mazzitelli
  (1997, bottom). (Right) A comparison of the cumulative distributions
which are used for the K-S tests described in the text. These model
  distributions are too narrow to adequately represent the data,
  implying a spread in true radius.
}
\label{gausszero}
\end{figure*}

\begin{table*}
\caption{A summary of the results obtained from comparing models with a
  dispersion in normalised radius (see text) with the observed
  distribution  of $R\sin i/R_{\rm 3Myr}$. The comparisons were done
  using either the S00 or DAM97 models to calculate $R_{\rm 3Myr}$.
  Columns 1--4 list the model parameters: the Gaussian sigma in $\log_{10}
  (R_{\rm true}/R_{\rm 3Myr})$, the threshold inclination for detection of
  periodic variability, the assumed $T_{\rm eff}$ uncertainty and the mean
  $\log_{10}$ age that best fits the data. Column 5
  lists the K-S statistic $D_{\rm max}$ (see Press et al. 1992),
  column 6 gives the integrated probability that $D>D_{\rm max}$, 
  corresponding to one minus the model rejection confidence level, and
  column 7 gives a comment on how well the model distribution matches
  the data.}
\begin{tabular}{ccccccl}
\hline
$\sigma_r$ & $i_{\rm th}$ & $\Delta T_{\rm eff}$ & Mean $\log_{10}$ &
$D_{\rm max}$ & $P(D>D_{\rm max})$ & comment \\
         & (deg)   & (K)   & Age/Myr  &   & &\\
\hline
&&&&&&\\
\multicolumn{7}{l}{Siess et al. (2000) models}\\
0.00 & 30 & 150 & 6.25 & 0.144 & 0.037 & too narrow\\
0.00 & 15 & 150 & 6.22 & 0.125 & 0.101 & marginally too narrow\\
0.00 & 30 & 300 & 6.25 & 0.101 & 0.275 & reasonable\\
0.05 & 30 & 150 & 6.26 & 0.124 & 0.103 & marginally too narrow\\
0.15 & 30 & 150 & 6.25 & 0.041 & 0.997 & good fit \\
0.27 & 30 & 150 & 6.26 & 0.124 & 0.103 & marginally too broad\\
&&&&&&\\
\multicolumn{7}{l}{D'Antona \& Mazzitelli (1997) models}\\
0.00 & 30 & 150 & 5.88 & 0.180 & 0.004 & too narrow\\
0.08 & 30 & 150 & 5.89 & 0.122 & 0.111 & marginally to narrow\\
0.15 & 30 & 150 & 5.88 & 0.060 & 0.876 & good fit\\
0.24 & 30 & 150 & 5.85 & 0.123 & 0.108 & marginally too broad \\
\hline
\end{tabular}
\label{results1}
\end{table*}

The first distribution of $R_{\rm true}$ tested was that of a coeval
population. The age was specified
and then the cumulative distribution of $(R\sin i)_{\rm obs}/R_{\rm
3Myr}$ tested against the Monte-Carlo predictions. The assumed age
was adjusted to minimise the K-S
statistic and select the model $R_{\rm true}$ distribution 
least different to the observational data. This process was repeated
using both the S00 and DAM97 evolutionary models to calculate $R_{\rm
3Myr}$.  The numerical results are listed in
Table~\ref{results1} and illustrated in Fig~\ref{gausszero} (the coeval
models are the rows with a dispersion in radius [see section 4.3],
set to $\sigma_{r}=0$). Initially
I used Gaussian $T_{\rm eff}$ uncertainties of 150\,K and a threshold
inclination $i_{\rm th}=30^{\circ}$. The assumed $v_{\rm true}$
distribution consists of 20 per cent of stars following a uniform
distribution for $10<v_{\rm true}<120$\,km\,s$^{-1}$ with the remaining
80 per cent following an exponential distribution with a decay constant
of 12\,km\,s$^{-1}$ between $10<v_{\rm
true}<120$\,km\,s$^{-1}$. Adopting this, the observed $v\sin i$
distribution is well reproduced by the model with a K-S rejection probability
of $<10$ per cent (see Jeffries 2007).

For both the DAM97 and S00 models the data are inconsistent with the
PMS population being drawn from a single coeval isochrone, at
confidence levels of 99.6 and 96.3 per cent respectively. 
These must be considered lower-limits to the model-rejection confidence
level because the mean age was adjusted as a free parameter to minimise
the K-S statistic.  The ``best-fitting'' ages are $\log_{10}$(age/Myr)
of 5.88 and 6.25 for the DAM97 and S00 models respectively. These are
similar to, but slightly older than, the median ages found from the H-R diagrams in
section~2, but note that values determined from the projected radii are
independent of the assumed distance to the ONC and unaffected by the
binary status of any of the stars.

Fig.~\ref{gausszero} shows that the coeval models are too narrow in $R
\sin i$.  Obviously a radius (or age) spread could broaden the model
distribution, but so too could a decrease in the assumed value of
$i_{\rm th}$ or an increase in the assumed $T_{\rm eff}$
uncertainties. To test this I ran the coeval S00 models assuming
$i_{\rm th}=15^{\circ}$. Table~\ref{results1} shows the ``best-fit''
parameters and that the coeval model is still rejected with $\simeq 90$
per cent confidence. Conversely, if $i_{\rm th}$ were greater than
$30^{\circ}$ a coeval model would be an even poorer match to the data.
I next tried broadening the model $R \sin i$ distribution by increasing
the uncertainty in $T_{\rm eff}$. The model can still be rejected at
$\geq 90$ per cent confidence unless the $T_{\rm eff}$ uncertainties
are approximately doubled to 300\,K, which is equivalent to a $\pm 2$
spectral subclass (1-sigma) uncertainty in the spectral types. Note
again that these confidence levels are lower limits because the mean
age was tuned to minimise the K-S statistic.

I conclude that unless the uncertainties in spectral classification and
$T_{\rm eff}$  have been significantly underestimated by
Hillenbrand (1997), then the observed values of $R \sin i$ are
inconsistent with coevality at a reasonably high level of confidence. 

\subsection{A radius dispersion}

\begin{figure*}
\centering
\begin{minipage}[t]{0.45\textwidth}
\includegraphics[width=80mm]{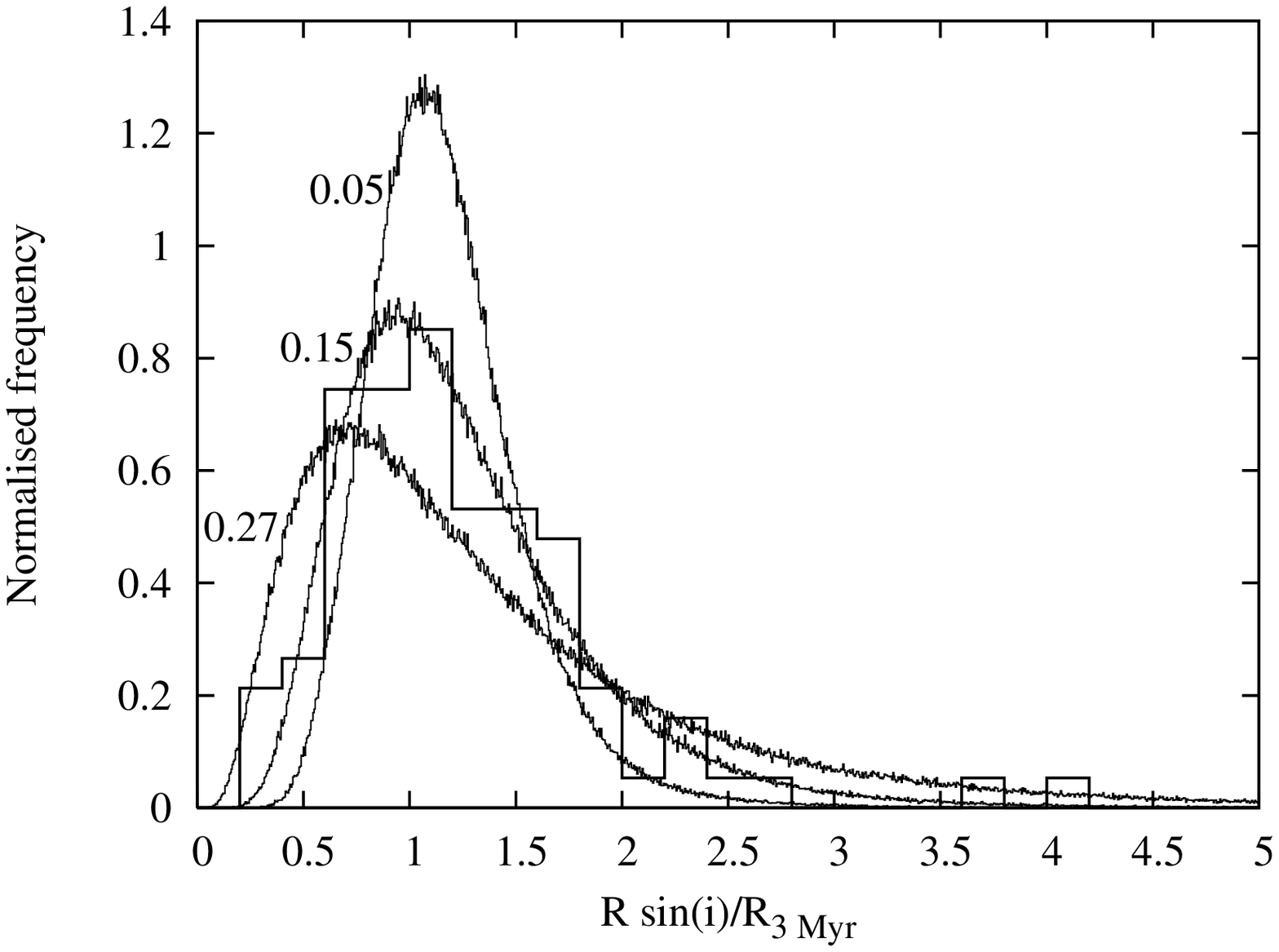}
\end{minipage}
\begin{minipage}[t]{0.45\textwidth}
\includegraphics[width=80mm]{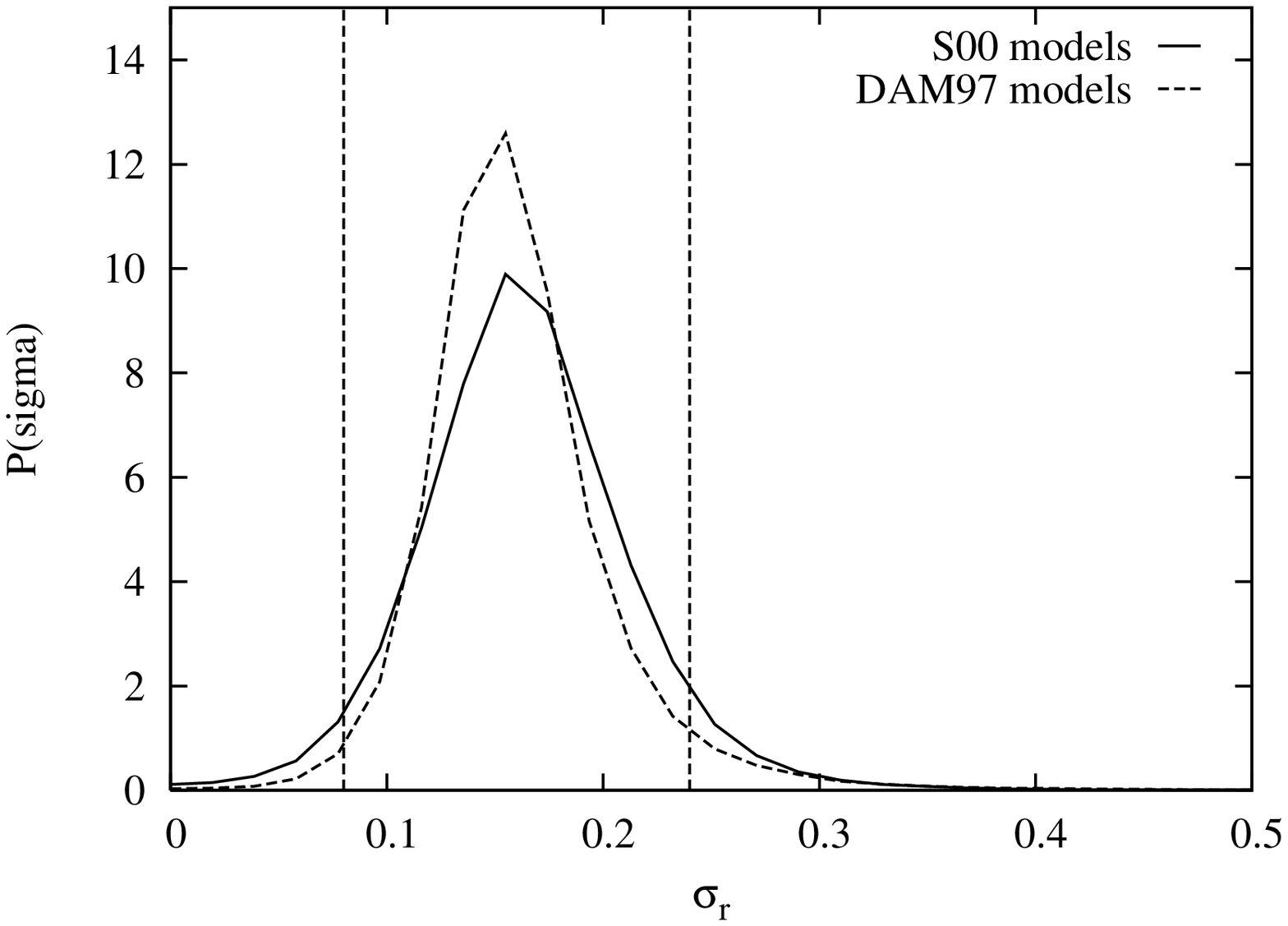}
\end{minipage}
\caption{The distribution of normalised $(R\sin i)_{\rm obs}$ for the
rotation sample compared with models that have a Gaussian distribution
in $\log_{10} R$ around an isochronal value.  The left hand panel shows
the comparison for Gaussian sigmas of 0.05, 0.15 and 0.27 respectively
-- corresponding to the best fit value and the values below or above
which the model can be rejected with $>90$ per cent
confidence. $R_{{\rm 3Myr}}$ was calculated using the S00 models.  The
right hand panel shows the normalised probability distribution,
integrated over all possible central $R_{\rm true}/R_{\rm 3Myr}$ values
(using either the S00 models [solid line] or the DAM97 models [dashed
line]), that these distributions are consistent with the observed data
as a function of the Gaussian dispersion, $\sigma_{r}$. The results
using both evolutionary models suggest a significant and similar spread
in radii.  The vertical dashed lines enclose 90 per cent of the
probability in the case of the S00 models.
}
\label{gaussr}
\end{figure*}

Using model~2 (a Gaussian distribution of $\log_{10} R$ about a central
value defined by a single isochronal age with a standard deviation of
$\sigma_r$), I simulated a dispersion in radius about an isochronal
value. The central isochronal age was again adjusted to minimise the
K-S statistic and a range of $\sigma_r$ was tested.  The results, using
the S00 models to calculate $R_{\rm 3Myr}$, are listed in
Table~\ref{results1} and illustrated in Fig.~\ref{gaussr}.  If the
dispersion increases to $\sigma_r>0.05$ then this model can no longer
be rejected with $>90$ per cent confidence. The ``best fit'' dispersion
corresponds to $\sigma_r=0.15$, whilst the model distribution becomes
too broad and can be rejected with $>90$ per cent confidence if
$\sigma_r>0.27$. The results using the DAM97 models are similar --
models with $0.08<\sigma_r<0.24$ cannot be rejected with better than 90
per cent confidence -- but the required central isochronal age is still
smaller ($\log_{10}$(age/Myr) is 5.88 compared with 6.25 when using the
S00 models -- see Table~\ref{results1}).

The results above cannot be used to estimate confidence intervals on
the best fitting value of $\sigma_r$, because the central isochronal age
has been optimised in each case. The right
hand panel of Fig.~\ref{gaussr} shows a normalised probability
distribution as a function of $\sigma_r$ (calculated using the S00
models), where the probability has been integrated over a wide range of
possible mean ages.  This illustrates that
the best fitting $\sigma_r$ is indeed about 0.15, with 90 per cent of the
probability distribution contained between 0.08 and 0.24. The results
using the DAM97 models are very similar
and give a slightly narrower 90 per cent confidence interval of $0.10 <
\sigma_r < 0.22$.

The Monte Carlo models suggest that the observed $R\sin i$ distribution
can be well represented if the observed population has a spread,
amounting to a full-width at half maximum of 0.2--0.5\,dex, in the
stellar radii around an isochronal value. This result is almost
independent of which evolutionary models are considered. An obvious
interpretation is that this is caused by a spread in ages and this is
investigated in the next subsection.

\subsection{An age spread}

\begin{figure*}
\centering
\begin{minipage}[t]{0.45\textwidth}
\includegraphics[width=80mm]{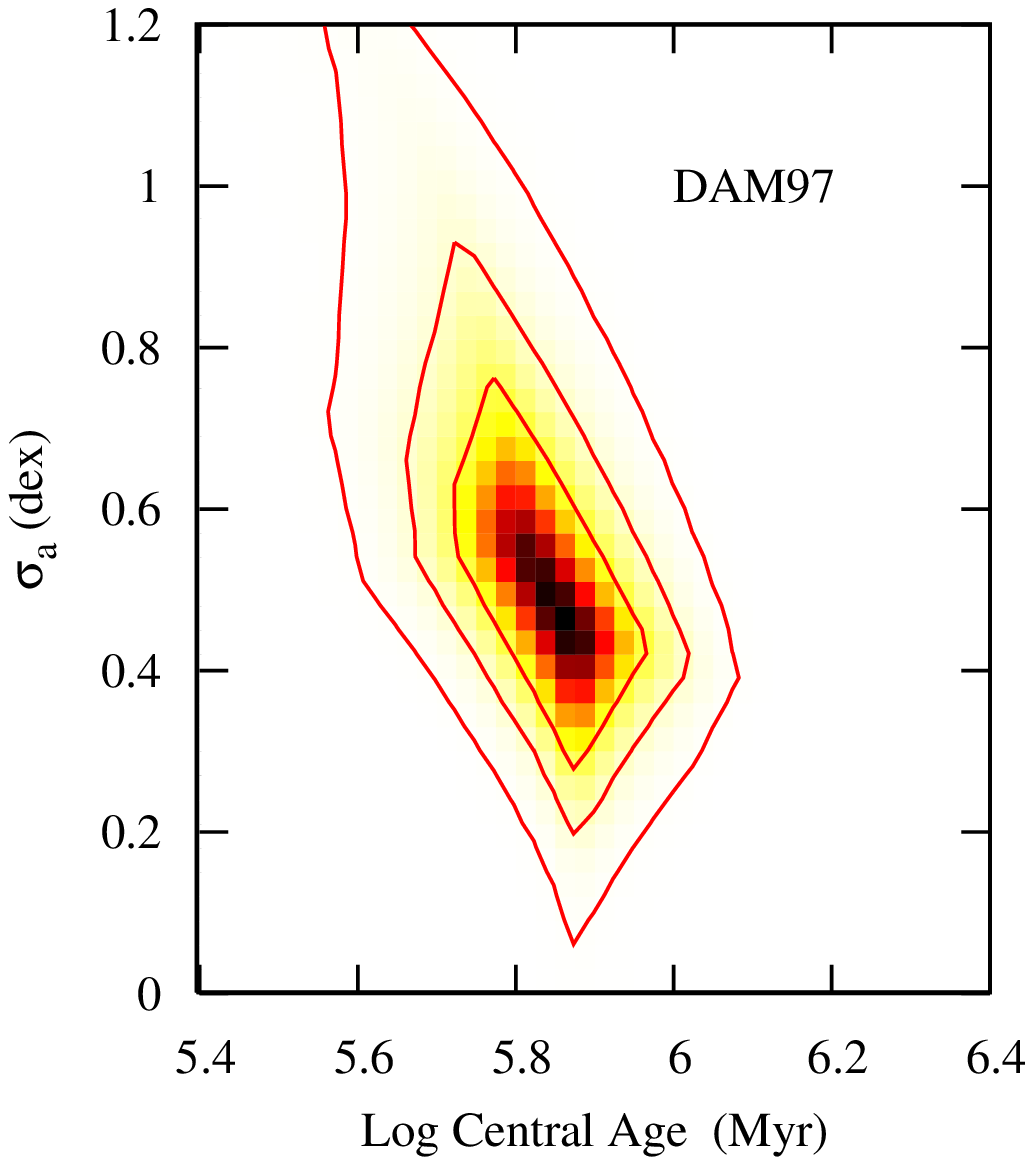}
\end{minipage}
\begin{minipage}[t]{0.45\textwidth}
\includegraphics[width=80mm]{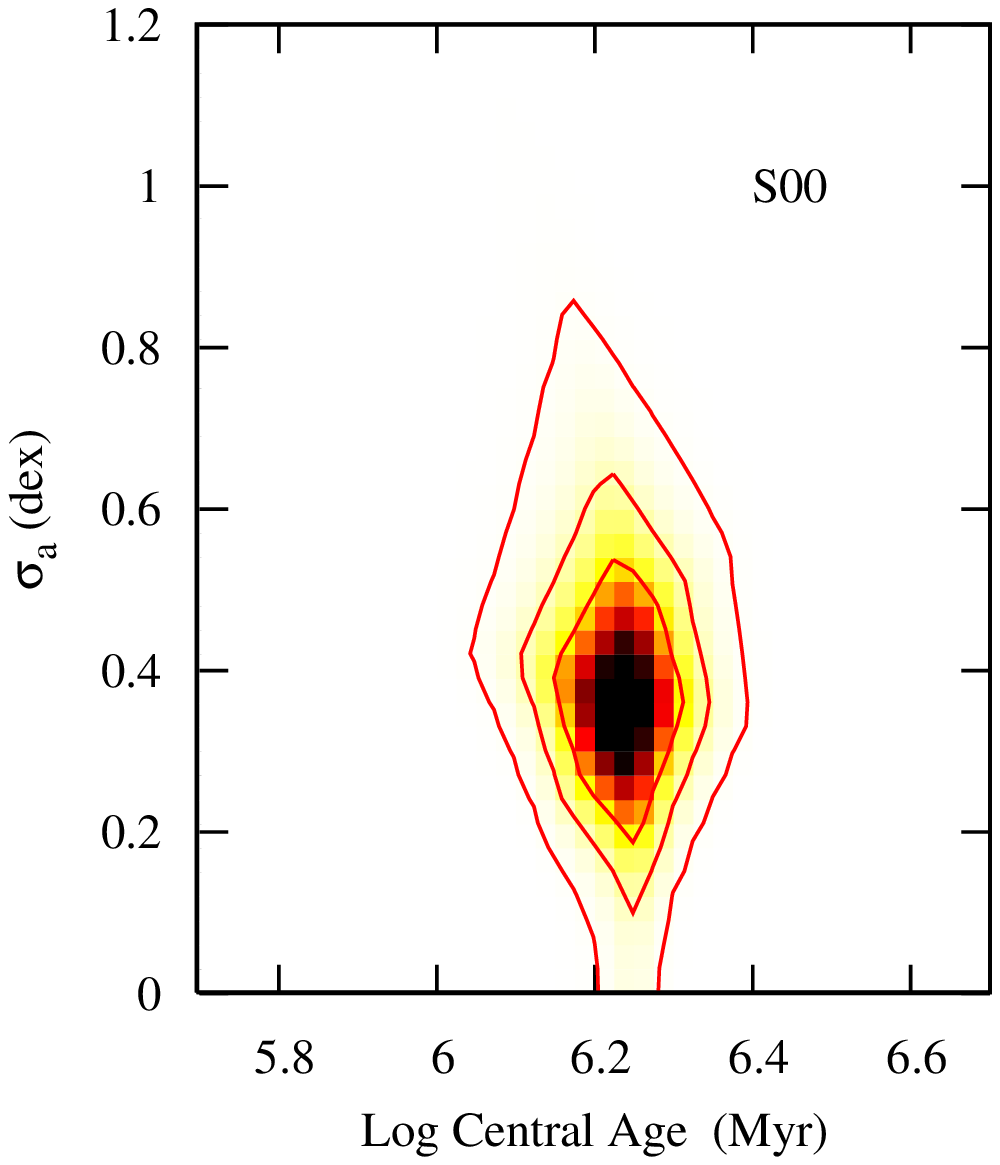}
\end{minipage}
\caption{Images representing the relative probability of a Gaussian
  distribution of $\log_{10}$ age being able to describe the observed
  distribution of $R\sin i$. The x-axes are the central isochronal age
  and the y-axes are the Gaussian dispersion ($\sigma_{a}$ in dex). The contours
  enclose 68, 90 and 99 per cent of the probability distribution.
  The left hand plot corresponds to radii calculated using the DAM97
  models; the right hand plot has radii calculated using the S00 models.
  Both plots have assumed $i_{\rm th}=30^{\circ}$ (see
  Table~\ref{results2}) and both suggest a spread in age that is larger
  than the median age of the cluster.
}
\label{pgrid}
\end{figure*}
\begin{figure*}
\centering
\begin{minipage}[t]{0.45\textwidth}
\includegraphics[width=80mm]{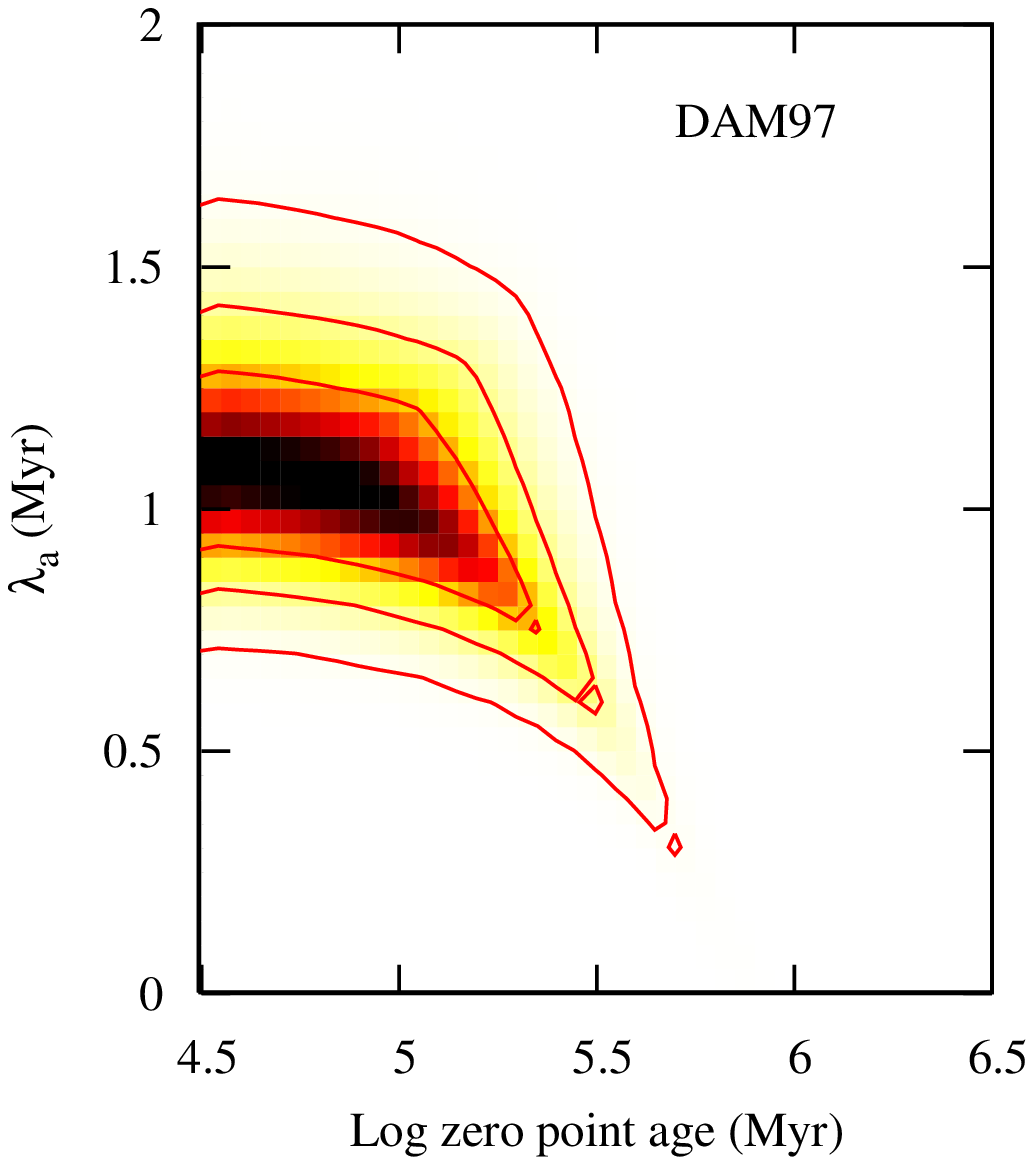}
\end{minipage}
\begin{minipage}[t]{0.45\textwidth}
\includegraphics[width=80mm]{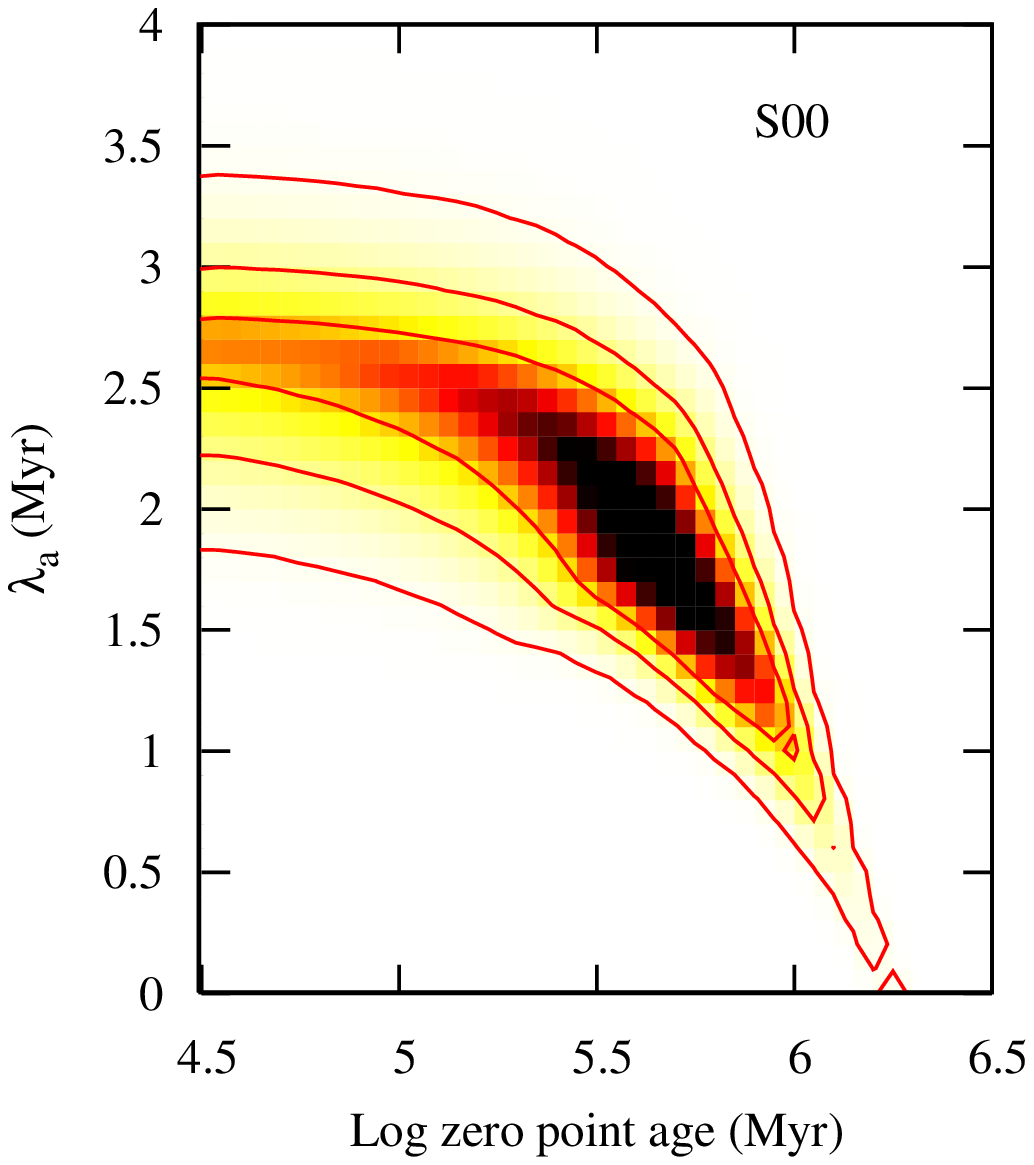}
\end{minipage}
\caption{Similar to Fig.~\ref{pgrid} but for an age distribution that
  decays exponentially with time scale $\lambda_{a}$ (on the y-axis) from a maximum at
  a zero point age (on the x-axis) and is zero prior to this. Left and
  right hand plots correspond to radii calculated using the DAM97 and
  S00 models. Contours enclose 68, 90 and 99 per cent of the relative
  probability distributions. In both case $\lambda_{a}$ is similar to
  the median cluster age using the same evolutionary models.
}
\label{pgrid2}
\end{figure*}

\begin{table*}
\caption{A summary of the results of modelling the observed
  distribution of 
$R\sin i/R_{\rm 3Myr}$ with two assumed age distributions (see text). For each
  assumed distribution results are given for the best-fitting cases where 
  radii were calculated using either the DAM97 or S00 evolutionary
  models and for two values of the threshold inclination angle. 
The final column gives the 90 percent confidence interval
  for $\sigma_{a}$ or $\lambda_{a}$ respectively.
}
\begin{tabular}{ccccccc}
\hline
\multicolumn{7}{l}{Gaussian dispersion in log age}\\
Model & $i_{\rm th}$ & Log central age/Myr & $\sigma_{a}$ & $D_{\rm max}$ & $P(D>D_{\rm
  max})$ & $\Delta \sigma_a$ \\
         &(deg) &        & (dex)   &       &     & (dex)  \\
DAM97    & 30   & 5.85  & 0.48 & 0.059 & 0.886 & 0.31--0.82 \\
DAM97    & 15   & 5.83  & 0.43 & 0.060 & 0.873 & 0.27--0.78 \\
S00      & 30   & 6.22  & 0.36 & 0.045 & 0.989 & 0.20--0.56 \\
S00      & 15   & 6.20  & 0.34 & 0.043 & 0.994 & 0.14--0.54 \\
&&&&&&\\
\multicolumn{7}{l}{Exponential age distribution}\\
Model &$i_{\rm th}$ &  Log zero point age/Myr & $\lambda_{a}$ & $D_{\rm max}$ & $P(D>D_{\rm
  max})$ & $\Delta \lambda_a$ \\
         & (deg)&        & (Myr)   &       &     & (Myr) \\
DAM97    & 30 & 4.55  & 1.10 & 0.058 & 0.898 & 0.72--1.30 \\
DAM97    & 15 & 4.95  & 0.95 & 0.057 & 0.911 & 0.58--1.15 \\
S00      & 30 & 5.65  & 1.90 & 0.043 & 0.993 & 1.22--2.82 \\
S00      & 15 & 5.65  & 1.70 & 0.037 & 0.999 & 0.98--2.59 \\            \\
\hline
\end{tabular}
\label{results2}
\end{table*}

Two separate models for an age spread were tested. The first was a
Gaussian distribution of $\log_{10}$ age around a central
isochrone. The free parameters were the central age and the age
dispersion (in logarithmic units), $\sigma_a$. The second model was an
age distribution which is zero up to some starting age, jumps to a
maximum and then decays exponentially with a decay constant,
$\lambda_a$, expressed in Myr. This latter model, with a suitably small
starting age, represents the exponentially accelerating star formation
model advocated by Palla \& Stahler (1999). The lowest starting age
considered was $0.03$\,Myr.

Ages drawn randomly from the Gaussian age distribution
were transformed into radii using the appropriate stellar
models (at the $T_{\rm eff}$ of each star in the observational
dataset) and these radii were perturbed according to the 150\,K $T_{\rm eff}$
uncertainties and then subjected to the measurement uncertainties,
random axial orientations and selection effects before comparing the
observed and modelled distribution of $R\sin i/R_{{\rm 3Myr}}$.
A grid of models covering a wide range of central
ages and age dispersion was calculated for both the DAM97 and S00 models.
Assuming that the ``correct'' solution lay within this grid, I
normalised the K-S probabilities and this gave a pair of relative probability
grids which are shown in Figs.~\ref{pgrid}a
and~\ref{pgrid}b. Projection of these grids onto the age dispersion
axis gave a best fitting $\sigma_{a}$ and a confidence interval
(analogous to the procedure used to create Fig.~\ref{gaussr}) which are
listed in Table~\ref{results2}. The same procedure was used for the
exponential age distribution, with the grid projected onto the
$\lambda_a$ axis. The normalised probability grids for this
distribution are shown in Figs.~\ref{pgrid2}a and~\ref{pgrid2}b and the
numerical results are also listed in Table~\ref{results2}.

There are acceptable fits to the observational data for both Gaussian
and exponential age distributions and for both the DAM97 and S00
models. Hence the current data are incapable of distinguishing between
these possibilities. As expected, the 90 per cent confidence
intervals on $\sigma_{a}$ and $\lambda_{a}$ do not encompass zero,
which would correspond to a coeval population. The dispersion in
$\log_{10}$ age for the Gaussian distribution is quite similar for the
DAM97 and S00 models: both models suggest a full-width half maximum
spread of factors of a few to $>10$ and therefore an overall spread in
absolute ages that is larger than the central age of the distribution and the
median age of the ONC found from the H-R diagram. The
exponential distribution has a favoured starting age of essentially
zero for the DAM97 models and a decay time scale, $\lambda_{a}$ of
1.1\,Myr. Again this indicates a spread of ages that is larger than the
median age of the sample from the H-R diagram. The S00 models favour a non-zero
maximum in the age distribution, which is also seen in the H-R
diagram-based age distribution shown in Fig.~\ref{age2}, although a
very young starting age cannot be ruled out. The decay timescale is
about 2\,Myr and again this implies a spread of ages that is larger
than the median age of the sample.

I ran some more simulations using a smaller value of $i_{\rm
th}=15^{\circ}$.  The results for these are also given in
Table~\ref{results2}. As expected a smaller $i_{\rm th}$ slightly
reduces the age dispersion required to explain the data, but does not
affect the broad conclusions. Conversely, a larger $i_{\rm th}$ would
increase the required age dispersion.

\subsection{Ages based on the Hertzsprung-Russell diagram}

\begin{figure*}
\centering
\begin{minipage}[t]{0.45\textwidth}
\includegraphics[width=80mm]{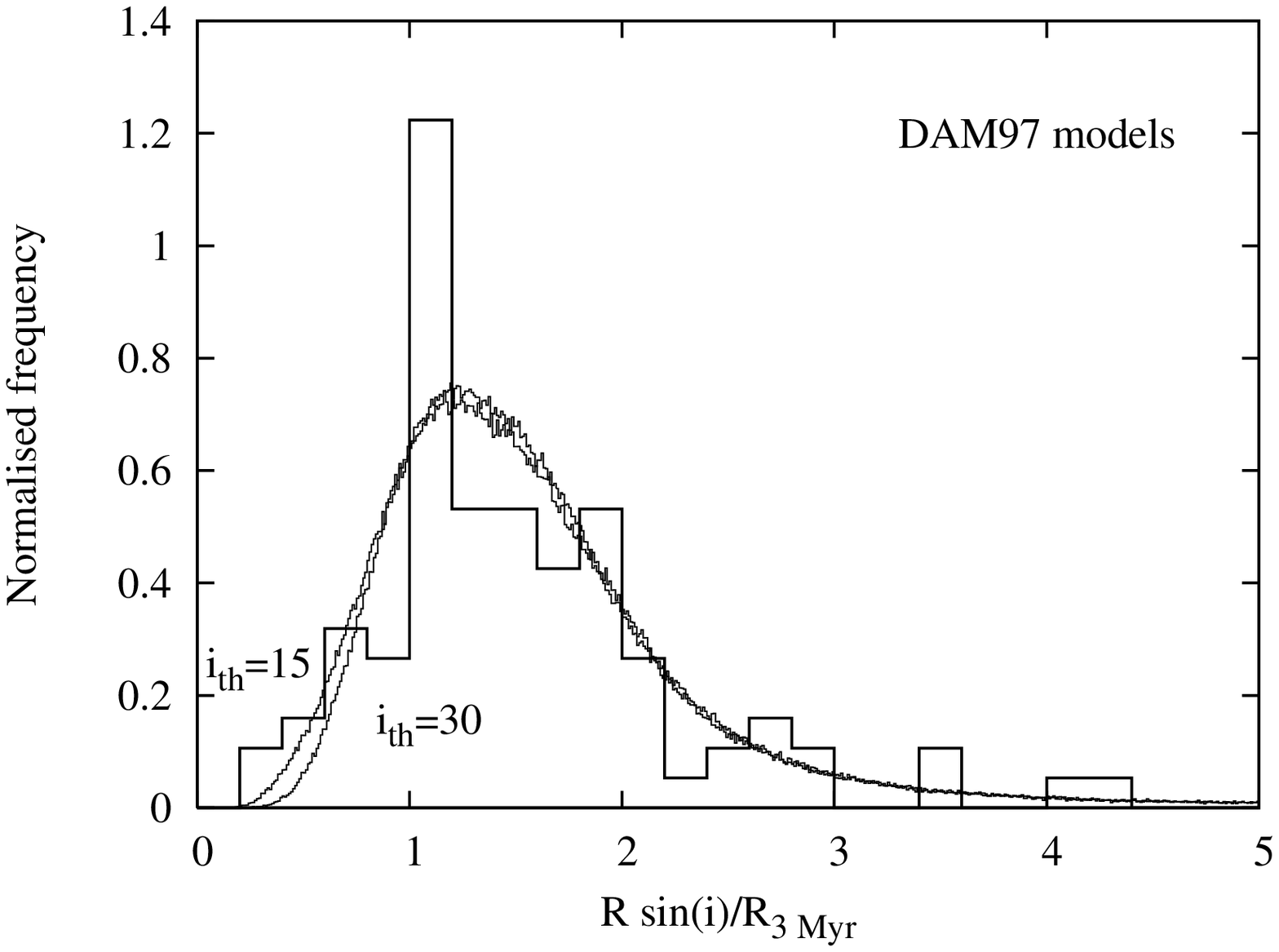}
\end{minipage}
\begin{minipage}[t]{0.45\textwidth}
\includegraphics[width=80mm]{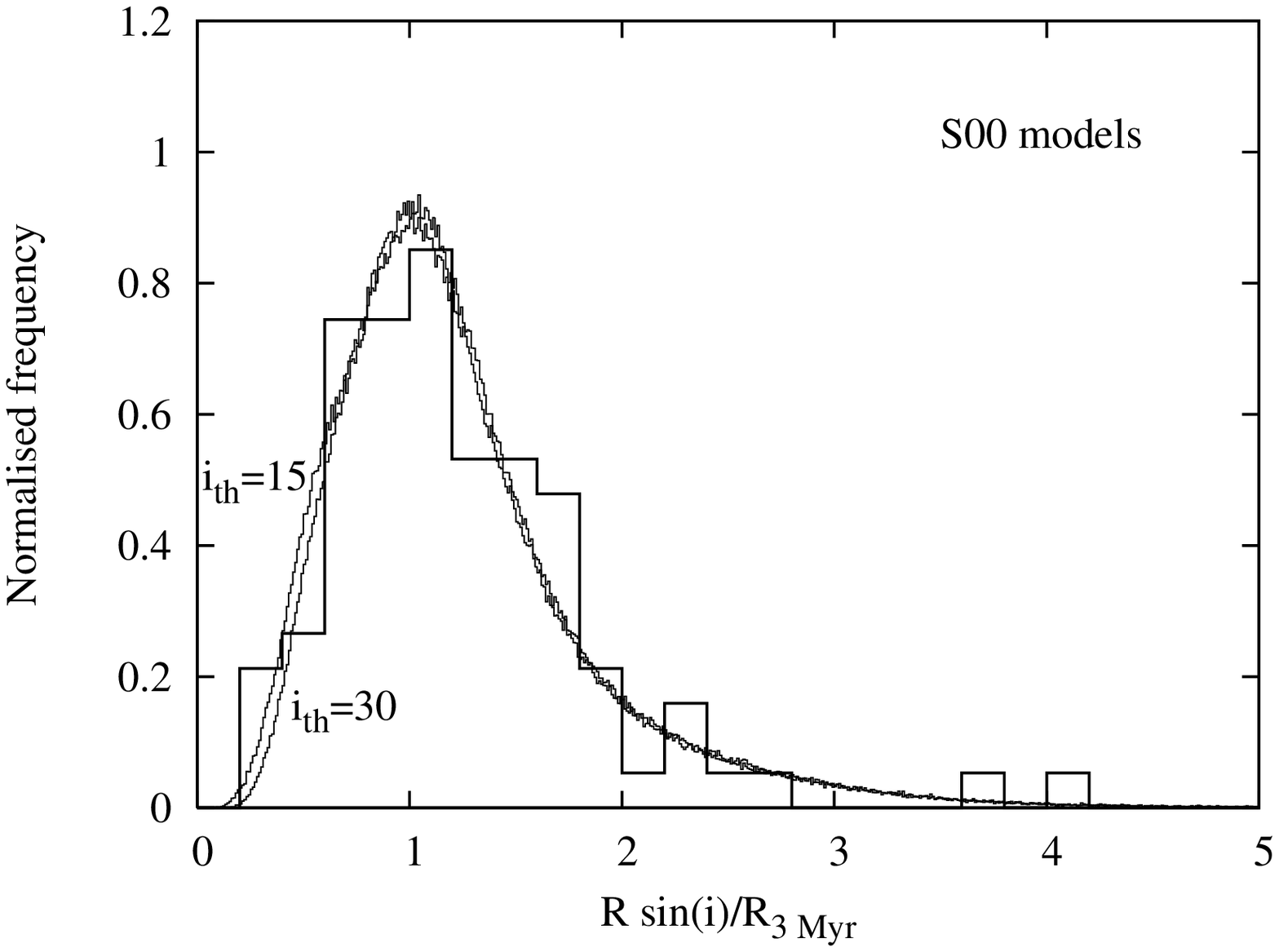}
\end{minipage}
\caption{Normalised projected radii distributions generated using the
  age distributions derived for the H-R diagrams in Fig.~\ref{age}
  and~\ref{age2}. The left and right hand panels show distributions using the
  DAM97 and S00 models respectively and that these match the observed
  distributions reasonably well. In each case models with $i_{\rm
  th}=15^{\circ}$ and $30^{\circ}$ are shown, demonstrating that this
  parameter does not affect the conclusions greatly.
}
\label{hragedist}
\end{figure*}

\begin{table}
\caption{A summary of the results of modelling the observed
  distribution of $R\sin i/R_{\rm 3Myr}$ using the age distribution
  determined from the Hertzspring-Russell diagram. The column headings
  are labelled as for Table 1.}
\begin{tabular}{ccccl}
\hline
Model & $i_{\rm th}$  &
$D_{\rm max}$ & $P(D>D_{\rm max})$ & comment\\
         & (deg)   &   & & \\
\hline
DAM97 & 30 & 0.119 & 0.130 & marginally narrow\\
DAM97 & 15 & 0.089 & 0.430 & good fit \\
S00   & 30 & 0.052 & 0.957 & good fit \\
S00   & 15 & 0.066 & 0.791 & good fit \\
\hline
\end{tabular}
\label{results3}
\end{table}

As a final test I constructed age distributions based upon the ages
determined from the H-R diagram for the rotation sample, which are
shown in the right hand panels of Figs.~\ref{age} and~\ref{age2}. These
age distributions were used to construct model $R\sin i/R_{\rm 3Myr}$
distributions which in turn were compared with the observed $R\sin
i/R_{\rm 3Myr}$ distributions via K-S tests. Initially, $i_{\rm th}$ was fixed at
$30^{\circ}$. A further set of models was generated with $i_{\rm
th}=15^{\circ}$ to investigate the influence of this parameter.

The comparison between the model and observed distributions of $R\sin
i/R_{\rm 3Myr}$ are shown in Fig.~\ref{hragedist}. Numerical results are
listed in Table~\ref{results3}. All of the model
distributions are formally compatible with the data and cannot be
rejected at the 90 per cent confidence level, although the DAM97
$i_{\rm th}=30^{\circ}$ model barely passes this test. Hence 
the age distributions determined from the
Hertzsprung-Russell diagram cannot be discounted as a valid description
of the age distribution required by the projected
radii. Given the uncertainties in determining ages from the H-R diagram
-- binarity, accretion, reddening, variability (see Hartmann 2001) --
that do not afflict determinations of the projected radii, this might
seem somewhat surprising. One might have thought that the age
distributions from the H-R diagram would be too broad because of these additional
sources of error.

The overall uncertainties in the luminosities have been considered by
a number of authors (e.g. Hillenbrand 1997; Hartmann 2001; Rebull et
al. 2004), who have concluded that Gaussian uncertainties in $\log_{10}$ luminosity
are in the range 0.16--0.20 dex. This leads approximately to Gaussian
uncertainties in deduced age of 0.24--0.30 dex for PMS stars (Hartmann
2001). In section 4.4 I found that the age dispersion required to
explain the normalised projected radius distribution was probably
larger than this. Hence it seems that the uncertainties that affect the
H-R diagram-based ages are not large enough to significantly broaden
the deduced age distribution beyond that which is indicated by the more
robust distribution of projected radii.

\section{Discussion}

Palla \& Stahler (1999) and Huff \& Stahler (2006) presented conventional
analyses of the ONC H-R diagram, concluding that star formation began at a
low level about 10\,Myr ago and accelerated up to the present
day. Slesnick et al. (2004) also found evidence for an older population
in the ONC H-R diagram they assembled, but argued it might be due to
younger stars which are viewed in scattered light from 
circumstellar material. Hartmann (2001) asserted that these large age
spreads could be an artefact of observational errors, binarity, variability,
differential reddening and accretion luminosity that conspire to cause
an observed spread in the H-R diagram that is not due to a
significant spread in age. Hartmann (2001) demonstrated that symmetric
uncertainties in $\log_{10}$ luminosity can lead to the appearance of
an exponentially decaying linear age distribution (see
Fig.~\ref{age}).  Palla et al. (2007) support their argument for a
true age spread by showing that a few percent of the ONC PMS population
show evidence of lithium depletion which is consistent with ages $\geq
10$\,Myr. 

The evidence presented in this paper supports the idea that at a given
$T_{\rm eff}$, the ONC stars in our sample do exhibit a spread in
radius (and hence luminosity) that is not consistent with isochronal
radii at a single age. One interpretation of this is that the ONC stars
have a range of ages. Our modelling shows that plausible age
distributions can be hypothesised which do explain the observed
dispersion in radii. These distributions have spreads in age which are
larger than the median age of the sample.

It is worth emphasising that the technique employed here is independent
of whether stars have correctly estimated reddening, whether they are
accreting, whether their distance is correctly estimated 
and other sources of uncertainty in calculating their
luminosities (see Hillenbrand 1997). The method does depend on
reasonable estimates of the measurement uncertainties in $v \sin i$ and
$T_{\rm eff}$ and the assumption that axial orientations are
random. The latter assumption is difficult to support or falsify at
present (see discussion in Jeffries 2007), although any concentration
of inclination angle would require a greater spread of intrinsic
radii or ages to match the observed spread in $R \sin i$.

A spread in ages is not the only way of
explaining a dispersion in radius at a given $T_{\rm eff}$. For
instance Tout, Livio \& Bonnell (1999) show that an ``age spread'' in
the H-R diagram could be produced in a coeval population of PMS stars
if they have very different accretion histories.  Irrespective of
whether the spread in radii found here is interpreted in terms of a
spread in age, there are important consequences for those using the ONC
(and other young star forming regions) to investigate evolutionary
trends of angular momentum loss, magnetic activity and circumstellar
discs (e.g. Herbst et al. 2005) and for those using
evolutionary models to estimate masses and find the (sub)stellar mass
function. In these cases, the often-used assumption that stars in a
given cluster are close-to-coeval or have similar properties
(luminosity, radius) at a given mass or $T_{\rm eff}$, is
not valid.

The sample examined in this paper is subject to a number of
limitations. Foremost among these is that I have only been able to
comment on the radius distribution of those stars for which periods and
$v \sin i$ measurements are available. Following the discussion in
section~2, it is clear that this ``rotation sample'' is deficient in
the small tail of stars which form the controversial elderly population
discussed by Palla \& Stahler (1999) and Palla et
al. (2007). 

Whilst the H-R diagram-based age distribution does a reasonable job of
reproducing the observed distribution of projected radii, it would be
unwise to extrapolate and conclude that the H-R diagram-based ages of
the apparently elderly population are reliable. The likely non-Gaussian
behaviour of some of the causes of uncertainty in PMS luminosities
(e.g. accretion disk scattering or variability) means that this small
fraction of outliers could yet be objects whose average intrinsic
luminosity has been severely underestimated. Fortunately, the
determination of periods is still quite feasible for the fainter ONC
members.  Large telescopes with multi-object, high resolution
spectroscopic capability should then be capable of determining $v \sin
i$ for some fraction of these. Even a small number of $R\sin i$ values
for these objects would be capable of revealing whether they have
smaller radii than the rest of the ONC population, as expected from
their relative positions in the H-R diagram.

\section{Summary}

I have taken a sample of PMS objects in the ONC with known rotation
periods and projected equatorial velocities, and modelled the resultant
distribution of projected equatorial radii with a simple Monte-Carlo
simulation. The simulation assumes that rotation axes are randomly
oriented in space and takes into account selection effects and
observational uncertainties in the data. A comparison between models
constructed using a variety of hypotheses concerning the intrinsic
distribution of stellar radii leads to the following conclusions.

\begin{enumerate}
\item The observed distribution of projected equatorial radii is
  inconsistent with the hypothesis that the ONC sample is coeval with a
  high level of confidence. This conclusion can only be weakened if
  uncertainties in spectral types and inferred effective temperatures
  are significantly larger than estimated in the original literature sources.

\item A spread in radius (assuming a Gaussian distribution in
  $\log_{10} R$) of 0.2--0.5 dex (full width half maximum) about
  an isochronal locus is required to match the data. This is almost
  independent of the choice of evolutionary model.

\item If the radius dispersion is modelled in terms of a distribution
  of ages, then either a Gaussian or exponentially decaying form can feasibly
  reproduce the data. The absolute values of the median age and age spread or decay
  timescales depend on which evolutionary models are used, but age
  spreads must be greater than the median age of the ONC.

\item The age distributions deduced from a conventional H-R diagram
  adequately reproduce the observed distribution of projected radii,
  independent of which evolutionary models are used. This suggests that
  uncertainties in deriving intrinsic luminosities are sufficiently
  small for the sample considered, that they do not mask the underlying
  spread in radii and inferred age.

\item The current data are biased in the sense that low-luminosity ONC
  members, and hence objects that are older in the H-R diagram, are not
  represented in published catalogues of projected equatorial velocity.
  This means that the radius and inferred age dispersions found here
  may be lower limits. In particular, the possibility that some cluster members
  are $\geq 10$\,Myr older than the bulk of the ONC population
  cannot be tested. However, given projected equatorial velocities for
  some of these objects it should easily be possible to check whether
  these objects have smaller radii consistent with their luminosity-based older ages.

\end{enumerate}

\nocite{hillenbrand97}
\nocite{hillenbrand98}
\nocite{herbst05}
\nocite{herbst02}
\nocite{slesnick04}
\nocite{huff06}
\nocite{dantona97}
\nocite{rhode01}
\nocite{sicilia05}
\nocite{cohen79}
\nocite{siess00}
\nocite{shu87}
\nocite{hartmannsfr01}
\nocite{tassis04}
\nocite{vazquez05}
\nocite{tan06}
\nocite{palla07}
\nocite{palla99}
\nocite{palla00}
\nocite{jeffriesoncdist07}
\nocite{sandstrom07}
\nocite{hartmann01}
\nocite{tout99}
\nocite{rebull04}
\nocite{press92}

\bibliographystyle{mn2e}  
\bibliography{iau_journals,master}


\bsp 

\label{lastpage}

\end{document}